\documentclass[11pt,twoside]{article}
\usepackage[ansinew]{inputenc}
\usepackage{amsthm}
\usepackage{amssymb,bm}
\usepackage{amsmath}
\usepackage[english]{babel}
\usepackage{array}
\usepackage{amsfonts}
\usepackage{latexsym}
\usepackage{longtable}
\numberwithin{figure}{section}
\numberwithin{table}{section}

\usepackage[pdftex]{color,graphicx}
\usepackage[usenames,dvipsnames,table]{xcolor}
\usepackage{float} 
\usepackage{subfigure} 
\usepackage{wrapfig} 
\usepackage[rflt]{floatflt} 
\graphicspath{{./Figures2NERebp/}}

\usepackage{multirow}
\date{}
\numberwithin{equation}{section}
\numberwithin{figure}{section}

\def \iN {I\!\!N}

\title{Empirical best prediction of small area bivariate parameters\footnote{Supported by the Instituto Galego de Estat\'{\i}stica,
by the grants MTM2017-82724-R and PGC2018-096840-B-I00 of the MINECO and by the Xunta de Galicia (Grupos de Referencia Competitiva ED431C-2016-015 and
Centro Singular de Investigaci\'on de Galicia ED431G/01), all of them through the ERDF.}}
\author{Mar\'{\i}a Dolores Esteban$^{1}$,  Mar\'{\i}a Jos\'e Lombard\'{\i}a$^{2}$, Esther L\'opez-Vizca\'{\i}no$^{3}$,\\ Domingo Morales$^{1}$, Agust\'{\i}n P\'erez$^{1}$
\vspace{0.1 cm}\\
{\small  $^{1}$Universidad Miguel Hern\'{a}ndez de Elche, Spain.}\\
{\small$^{2}$Universidade da Coru\~na, CITIC, Spain,}\\
{\small $^{3}$Instituto Galego de Estat\'{\i}stica, Spain},\\
{\small July 16, 2020}}

\begin{document}
\maketitle
\begin{abstract}
This paper introduces empirical best predictors of small area bivariate parameters, like ratios of sums or sums of ratios,
by assuming that the target unit-level vector follows a bivariate nested error regression model.
The corresponding means squared errors are estimated by parametric bootstrap.
Several simulation experiments empirically study the behavior of the introduced statistical methodology.
An application to real data from the Spanish household budget survey gives estimators of ratios of food household expenditures by
provinces.
\vspace{0.1 cm}
\end{abstract}

\section{Introduction}
Complex indicators based on more than one variable play an important role in public statistics.
For a finite population, partitioned in domains or small areas, examples of such indicators are the ratios of domain means or the domain means of ratios.
In the first case, we may have the quotient between the average annual expenditure on food of the households from a given territory and
the corresponding average annual expenditure on all items of expenditure.
In the second case, we have the domain mean of the households ratios of food expenditure.

One way to estimate a ratio of domain means is to estimate the numerator and denominator separately and independently and substitute in its expression.
This approach leads to the use of plug-in estimators, which have the problem of being biased even though their components are unbiasedly estimated.
There are two additional inconveniences.
The first one is not considering the correlation between the variables that intervene in the definition of the population parameters of the ratio type.
The second one is that the asymptotic property of unbiasedness cannot be assumed for estimators of domain indicators if sample sizes are small.

For estimating domain means of households ratios of food expenditure, as well as for other nonlinear bivariate parameters,
we have not found any model-based estimators in the statistical literature. This paper partially covers this gap.

Small area estimation (SAE) gives statistical methodology to estimate parameters of population subsets, called domains or small areas.
The word ``small'' refers to sample size and not to population size.
To overcome the problem of having a small sample size in a domain, SAE complements the data of the target variable with data of auxiliary variables,
information taken from other domains and correlation structures.
All this can be done by fitting models to the available data for the entire population and building estimators based on the selected model.
This is the unit-level model-based approach.
Alternatively, models can be used for aggregated data and then inferential procedures are based on area-level models.

The domain parameters are functions of the values that one or several objective variables take in all the units of the population.
The mathematical expression (formula) of the domain parameter is therefore relevant.
If the target variables are continuous, then it is possible to estimate linear parameters with empirical best linear unbiased predictors based on linear mixed models (LMM).
However, domain parameters are often nonlinear or defined by non-continuous target variables.
In those cases, it is quite common to estimate domain parameters with empirical best (or Bayes) predictors (EBP) based on LMMs or on generalized LMMs (GLMM).
This paper deals with the estimation of domain parameters that are nonlinear functions of two continuous variables and
puts special emphasis in the estimation of ratios of domain means and domain means of ratios.

As the domain parameters of interest depend on several target variables, the use of multivariate models is recommended.
Since the first works of Fay (1987), Datta et al. (1991) and Datta et al. (1999), the statistical literature contains some applications of these models to the SAE setup.
Concerning area-level multivariate models,
Molina et al. (2007), L\'opez-Vizca\'{\i}no et al. (2013, 2015) and Esteban et al. (2020) derived predictors for totals of employed and unemployed people and for unemployment rates
based on multinomial-logit or compositional mixed models.
Morales et al. (2015), Porter et al. (2015), Benavent et al. (2016) or  Arima et al. (2017) studied the problem of estimating poverty indicators, including the non linear poverty gap.
Marchetti and Secondi (2017) Ubaidillah et al. (2019) estimated household consumption expenditures by applying Fay-Herriot models.
Concerning unit-level multivariate models,
Ngaruye (2017), Ito and Kubosawa (2018) and Esteban et al. (2021) gave BLUPs of domain means and totals to treat problems of repeated measures, posted land prices and expenditure data respectively.

On the other hand, as the domain parameters of interest are non linear then the use of EBPs are widely employed.
Since the first works of Jiang and Lahiri (2001) and Jiang (2003), where EBPs of functions of fixed effects and small-area-specific random effects were developed under GLMMs,
some authors have extended their procedures and applied EBPs in the SAE context.
For example, Boubeta et al. (2016, 2017) and Hobza et al. (2016, 2018) derived EBPs of small area poverty proportions based on
area-level Poisson mixed models and unit-level logit mixed models respectively.
Marino et al. (2019) and Hobza et al. (2020) proposed EBPs under semiparametric and parametric unit-level GLMMs.
Chandra et al. (2017, 2018) and Chandra, and Salvati (2018) introduced SAE predictors of spatially correlated count data.
Torabi (2019) proposed a class of spatial GLMMs to obtain small area predictors of esophageal cancer prevalence.
This uncomplete list of contribution shows the high impact that the EBP approach has in SAE.

Based on the nested error regression (NER) model, the  seminal paper of Molina and Rao (2010) introduced the basic theory for
calculating EBPs of domain nonlinear parameters, depending on one continuous target variable, under unit-level LMMs.
Their results where later extended to the two-fold NER model by Marhuenda et al. (2017), to lognormal models by Molina and Mart\'{\i}n (2018) and to
Data-driven transformations by Rojas-Perilla et al. (2020).
However, the statistical literature has not yet treated the problem of constructing EBPs, based on bivariate NER models,
for estimating domain nonlinear parameters defined by two target continuous variables.
This is the main contribution of this paper.

By following Gonz\'alez-Manteiga et al. (2007, 2008), this manuscript introduces a parametric bootstrap procedure for estimating the mean squared errors (MSE) of the EBPs.
As the optimality properties of the best predictors might not hold for EBPs when number of domains and domain sample sizes are small, an empirical research is carried out.
In addition to the mathematical developments, Monte Carlo simulations empirically investigates the properties of the EBPs and the corresponding MSE estimators.
Finally, the new statistical methodology is applied to data from the 2016 Spanish family budget survey.
The target is the estimation means of ratios and ratios of means of food expenditures in Spanish households at the province level.

The rest of the paper is organized as follows.
Section \ref{sect.2NER} introduces the bivariate nested error regression model.
Section \ref{sect.EBP} derives the EBPs of non-linear and ratio-type predictors.
Section \ref{sect.boot} describes a parametric bootstrap procedure to estimate MSEs of the introduced predictors.
Section \ref{sect.sim} carries out simulation experiments to investigate the behavior of the predictors of domain ratios and the MSE estimators.
Section \ref{sect.app} gives an application to data from the Spanish household budget survey of 2016,
where the target is the small area estimation of means of ratios and ratios of means of household annual food expenditures by provinces.
Finally,  Section \ref{sect.con} summarizes some conclusions.
\section{The model}\label{sect.2NER}
%
%
%
%
%
%
Let $U$ be a population of size $N$ partitioned into $D$ domains or areas $U_1,\ldots,U_D$ of sizes $N_1,\ldots,N_D$ respectively.
Let $N=\sum_{j=1}^DN_d$ be the global population size.
Let $y_{dj}=(y_{dj1},y_{dj2})^{\prime}$ be a vector of continuous variables measured on the sample unit $j$ of domain $d$, $d=1,\ldots,D$, $j=1,\ldots,N_d$.
For $k=1,2$, let ${x}_{djk}=(x_{djk1},\ldots,x_{djkp_k})$ be a row vector containing $p_k$ explanatory variables and let
${X}_{dj}=\mbox{diag}\left({x}_{dj1},{x}_{dj2}\right)_{2\times p}$ with $p=p_1+p_2$.
Let $\beta_{k}$ be a column vector of size $p_k$  containing regression parameters and let
$\beta=\left(\beta_{1}^{\prime},\beta_{2}^{\prime}\right)^{\prime}_{p\times1}$.
The population bivariate NER (BNER) model assumes that
\begin{equation}\label{modpop2NERdj}
y_{dj}=X_{dj}\beta+u_d+e_{dj},\quad d=1,\ldots,D,\, j=1,\ldots,N_d.
\end{equation}
where the vectors of random effects $\{u_{d}\}$ and random errors $\{e_{dj}\}$ are independent with multivariate distributions
$$
u_{d}\sim N_2(0,V_{ud}),\quad
e_{dj}\sim N_2(0,V_{edj})
$$
and covariance matrices
$$
{V}_{ud}=
\left(\begin{array}{cc}
\sigma_{u1}^2 & \rho_{u12}\sigma_{u1}\sigma_{u2} \\
\rho_{u12}\sigma_{u1}\sigma_{u2} & \sigma_{u2}^2 \\
\end{array}\right),\quad
{V}_{ud}=
\left(\begin{array}{cc}
\sigma_{u1}^2 & \rho_{u12}\sigma_{u1}\sigma_{u2} \\
\rho_{u12}\sigma_{u1}\sigma_{u2} & \sigma_{u2}^2 \\
\end{array}\right),
$$
with parameters $\theta_{1}=\sigma_{u1}^2$, $\theta_{2}=\sigma_{u2}^2$, $\theta_{3}=\rho_{u12}$, $\theta_{4}=\sigma_{e1}^2$, $\theta_{5}=\sigma_{e2}^2$ and $\theta_{6}=\rho_{e12}$.
Let ${I}_m$ be the $m\times m$ identity matrix. We define the $2N_d\times 1$ vectors and the $2N_d\times p$ and $2N_d\times 2$ matrices
$$
y_d=\underset{1\leq j \leq N_d}{\hbox{col}}(y_{dj}),\,\,\,
e_d=\underset{1\leq j \leq N_d}{\hbox{col}}(e_{dj}),\,\,\,
X_d=\underset{1\leq j \leq N_d}{\hbox{col}}(X_{dj}),\,\,\,
Z_d=\underset{1\leq j \leq N_d}{\hbox{col}}(I_2).
$$
Model (\ref{modpop2NERdj}) can be written in the domain-level form
\begin{equation}\label{modpop2NERd}
y_{d}=X_{d}\beta+Z_du_d+e_{d},\quad d=1,\ldots,D,
\end{equation}
where $u_d\sim N_2(0,{V}_{ud})$, $e_d\sim N_{2N_d}(0,V_{ed})$ are independent and $V_{ed}=\underset{1\leq j \leq N_d}{\hbox{diag}}(V_{edj})$.
We define the $2N\times 1$ and $2D\times 1$ vectors and the $2N\times p$ and $2N\times 2D$ matrices
$$
y=\underset{1\leq d \leq D}{\hbox{col}}(y_{d}),\,\,\,
e=\underset{1\leq d \leq D}{\hbox{col}}(e_{d}),\,\,\,
u=\underset{1\leq d \leq D}{\hbox{col}}(u_{d}),\,\,\,
X=\underset{1\leq d \leq D}{\hbox{col}}(X_{d}),\,\,\,
Z=\underset{1\leq d \leq D}{\hbox{diag}}(Z_d).
$$
Model (\ref{modpop2NERdj}) can be written in the linear mixed model form
\begin{equation}\label{modpop2NER}
y=X\beta+Zu+e.
\end{equation}
where $u\sim N_{2D}(0,{V}_{u})$, $e\sim N_{2N}(0,V_{ed})$ are independent,
$V_{u}=\underset{1\leq d \leq D}{\hbox{diag}}(V_{ud})$ and
$V_{e}=\underset{1\leq d \leq D}{\hbox{diag}}(V_{ed})$.

In practice, inference is carried out based on a sample, $s=\cup_{d=1}^Ds_d$, drawn from the finite population $U$.
Let $y_s$ and $y_{ds}$ be the sub-vectors of $y$ and $y_d$ corresponding to sample elements and $y_r$ and $y_{dr}$ the sub-vectors of $y$ and $y_d$ corresponding to the out-of-sample elements.
Without lack of generality, we can write $y_d=(y_{ds}^\prime,y_{dr}^\prime)^\prime$.
Define also the corresponding decompositions of $X_d$, $Z_d$ and $V_d$.

We assume that sample indexes are fixed, so that the sample sub-vectors $y_{ds}$ follow the marginal models derived from the population model (\ref{modpop2NERd}), i.e.
$$
y_{ds}=X_{ds}\beta+Z_{ds}u_d+e_{ds},\quad d=1,\ldots,D,
$$
where $u_d\sim N_2(0,{V}_{ud})$, $e_{ds}\sim N_{2n_d}(0,V_{eds})$ are independent and $V_{eds}=\underset{1\leq j \leq n_d}{\hbox{diag}}(V_{edj})$.
The vectors $y_{ds}$ are independent with $y_{ds}\sim N_{n_d}(\mu_{ds},V_{ds})$, $\mu_{ds}=X_{ds}\beta$, $V_{ds}=Z_{ds}V_{ud}Z_{ds}^\prime+V_{eds}$.
When the variance component parameters are known, the best linear unbiased estimator (BLUE) of $\beta$ and the best linear unbiased predictor (BLUP) of $u_d$, $d=1,\ldots,D$, are
\begin{equation}\label{BLUP}
\hat{\beta}_B=(X_s^{\prime}V_s^{-1}X_s)^{-1}X_s^{\prime}V_s^{-1}y_s,\quad
\hat{u}_{Bd}=V_{ud}Z_{ds}^\prime V_{ds}^{-1}\big(y_{ds}-X_{ds}\hat{\beta}_B\big).
\end{equation}
Esteban et al. (2021) introduces a Fisher-scoring algorithm for estimating the residual maximum likelihood estimators (REML) of the model parameters.
By substituting parameters by REML estimators in (\ref{BLUP}), the empirical BLUE and BLUP are obtained.

%
%
%
%
%
%
The non-sampled sub-vectors $y_{dr}$ follow the marginal models derived from the population model (\ref{modpop2NERd}), i.e.
$$
y_{dr}=X_{dr}\beta+Z_{dr}u_d+e_{dr},\quad d=1,\ldots,D,
$$
where $u_d\sim N_2(0,{V}_{ud})$, $e_{dr}\sim N_{2(N_d-n_d)}(0,V_{eds})$ are independent and $V_{edr}=\underset{n_d+1\leq j \leq N_d}{\hbox{diag}}(V_{edj})$.
The vectors $y_{dr}$ are independent with $y_{dr}\sim N_{N_d-n_d}(\mu_{dr},V_{dr})$, $\mu_{dr}=X_{dr}\beta$, $V_{dr}=Z_{dr}V_{ud}Z_{dr}^\prime+V_{edr}$.
The covariance matrix between $y_{dr}$ and $y_{ds}$ is
$$
V_{drs}=\mbox{cov}(y_{dr},y_{ds})=
\mbox{cov}(X_{dr}\beta+Z_{dr}u_d+e_{dr}, X_{ds}\beta+Z_{ds}u_d+e_{ds})
=Z_{dr}\mbox{var}(u_d)Z_{ds}^\prime=
Z_{dr}V_{ud}Z_{ds}^\prime.
$$
The distribution of $y_{dr}$, given the sample data $y_{s}$, is
\begin{equation}\label{conditioned}
y_{dr}|y_{s}\sim y_{dr}|y_{ds}\sim N(\mu_{dr|s},V_{dr|s}).
\end{equation}
The conditional $(N_d-n_d)\times1$ mean vector is
\begin{eqnarray*}
\mu_{dr|s}&=&\mu_{dr}+V_{drs}V_{ds}^{-1}(y_{ds}-\mu_{ds})=
X_{dr}\beta+Z_{dr}V_{ud}Z_{ds}^\prime V_{ds}^{-1}(y_{ds}-X_{ds}\beta)
\\
&=&X_{dr}\beta+Z_{dr}V_{ud}Z_{ds}^\prime \Big\{V_{eds}^{-1}-V_{eds}^{-1}Z_{ds}\Big(V_{ud}^{-1}+{n_d}V_{edj}^{-1}\Big)^{-1}Z_{ds}^\prime V_{eds}^{-1}\Big\}(y_{ds}-X_{ds}\beta).
\end{eqnarray*}
The conditional covariance matrix is
\begin{eqnarray*}
V_{dr|s}&=&V_{dr}-V_{drs}V_{ds}^{-1}V_{dsr}
=
Z_{dr}V_{ud}Z_{dr}^\prime+V_{edr}
-Z_{dr}V_{ud}Z_{ds}^\prime V_{ds}^{-1}Z_{ds}V_{ud}Z_{dr}^\prime
\\
&=&Z_{dr}V_{ud}Z_{dr}^\prime+V_{edr}
-Z_{dr}V_{ud}Z_{ds}^\prime
\Big\{V_{eds}^{-1}-V_{eds}^{-1}Z_{ds}\Big(V_{ud}^{-1}+{n_d}V_{edj}^{-1}\Big)^{-1}Z_{ds}^\prime V_{eds}^{-1}\Big\}
Z_{ds}V_{ud}Z_{dr}^\prime
\\
&=&Z_{dr}V_{ud}Z_{dr}^\prime+V_{edr}-{n_d}Z_{dr}V_{ud}V_{edj}^{-1}V_{ud}Z_{dr}^\prime
+n_d^2Z_{dr}V_{ud}V_{edj}^{-1}\Big(V_{ud}^{-1}+{n_d}V_{edj}^{-1}\Big)^{-1}V_{edj}^{-1}V_{ud}Z_{dr}^\prime.
\end{eqnarray*}
If $n_d\neq0$ and $j\in r_d=U_d-s_d$, $j>n_d$, the conditional $2\times1$ mean vector is
\begin{eqnarray*}
\mu_{dj|s}&=&X_{dj}\beta+V_{ud}Z_{ds}^\prime \Big\{V_{eds}^{-1}-V_{eds}^{-1}Z_{ds}\big(V_{ud}^{-1}+n_dV_{edj}^{-1}\big)^{-1}Z_{ds}^\prime V_{eds}^{-1}\Big\}(y_{ds}-X_{ds}\beta)
\\
&=&X_{dj}\beta+V_{ud}\Big\{I_2-n_dV_{edj}^{-1}\big(V_{ud}^{-1}+n_dV_{edj}^{-1}\big)^{-1}\Big\}\sum_{j=1}^{n_d}V_{edj}^{-1}(y_{dj}-X_{dj}\beta).
\end{eqnarray*}
If $n_d=0$ and $j\in r_d$, the conditional $2\times1$ mean vector is
$$
\mu_{dj|s}=X_{dj}\beta.
$$
If $n_d\neq0$ and $j\in r_d$, $j>n_d$, the conditional $2\times2$ covariance matrix is
$$
V_{dj|s}=V_{d|s}=V_{ud}+V_{edj}-n_dV_{ud}V_{edj}^{-1}V_{ud}
+n_d^2V_{ud}V_{edj}^{-1}\Big(V_{ud}^{-1}+n_dV_{edj}^{-1}\Big)^{-1}V_{edj}^{-1}V_{ud}.
$$
If $n_d=0$ and $j\in r_d$, the conditional $2\times2$ covariance matrix is
$$
V_{dj|s}=V_{d|s}=V_{ud}+V_{edj}.
$$
\section{EBPs of domain parameters}\label{sect.EBP}
\subsection{EBPs of additive parameters}\label{sect.ebp2NERaditive}
Let $z_{dj}=(z_{dj1},z_{dj2})^{\prime}$ be a vector of continuous positive variables measured on the sample unit $j$ of domain $d$, $d=1,\ldots,D$, $j=1,\ldots,n_d$.
This section consider additive domain $2\times1$ or $1\times 1$ parameters that can be written in the form
\begin{equation}\label{addparam}
\delta_d=\frac{1}{N_d}\sum_{j=1}^{N_{d}}h(z_{dj}),\quad d=1,\ldots,D,
\end{equation}
where $h$ is a known measurable function $R^2\mapsto R^t$, $t=1,2$.
Examples of real-valued function $h:R^2\mapsto R$ are $h(z_{dj})=z_{dj1}$, $h(z_{dj})=z_{dj2}$, and $h(z_{dj})=z_{dj1}/(z_{dj1}+z_{dj2})$.
The corresponding domain parameters (averages of marginal variables or of unit-level ratios) are
\begin{equation}\label{averageratio}
\overline{Z}_{d1}=\frac{1}{N_d}\sum_{j=1}^{N_d}z_{dj1},\quad
\overline{Z}_{d2}=\frac{1}{N_d}\sum_{j=1}^{N_d}z_{dj2},\quad
A_d = \frac{1}{N_d}\sum_{j=1}^{N_d}\frac{z_{dj1}}{z_{dj1}+z_{dj2}},\quad d=1,\ldots,D.
\end{equation}
In applications to real data $z_{dj1}$ and $z_{dj2}$ might not follow normal distributions, as may happens with expenditure variables that are typically asymmetric.
This is why we assume that there exist a one-to-one transformation $g:R^2\mapsto R^2$ such that $y_{dj}=g(z_{dj})$ follows the BNER model (\ref{modpop2NERdj}).
We further assume that $g$ is separable, i.e.
$$
y_{dj}=g(z_{dj})=\big(g_1(z_{dj1}),g_2(z_{dj2})\big)^\prime,\quad
z_{dj}=g^{-1}(y_{dj})=\big(g_1^{-1}(y_{dj1}),g_2^{-1}(y_{dj2})\big)^\prime,
$$
where $g_1:(0,\infty)\mapsto R$ and $g_2:(0,\infty)\mapsto R$ are one-to-one functions.
For $d=1,\ldots,D$, we write (\ref{addparam}) and (\ref{averageratio}) as functions of $y_{dj1}$ and $y_{dj2}$, i.e.
\begin{eqnarray*}
\delta_d&=&\frac{1}{N_d}\sum_{j=1}^{N_{d}}h(g^{-1}(y_{dj})),\quad
\overline{Z}_{d1}=\frac{1}{N_d}\sum_{j=1}^{N_d}g^{-1}_1(y_{dj1}),
\\
\overline{Z}_{d2}&=&\frac{1}{N_d}\sum_{j=1}^{N_d}g^{-1}_1(y_{dj2}),\quad
A_d = \frac{1}{N_d}\sum_{j=1}^{N_d}\frac{g^{-1}_1(y_{dj1})}{g^{-1}_1(y_{dj1})+g^{-1}_2(y_{dj2})}.
\end{eqnarray*}
The best predictor (BP) of $\delta_d$ is
\begin{eqnarray*}
\hat\delta_d^B&=&E_{y_r}\Big[\frac{1}{N_d}\sum_{j=1}^{N_{d}}h(g^{-1}(y_{dj}))\big|y_s\Big]
\\
&=&
\frac{1}{N_d}\Big\{\sum_{j\in s_{d}}h(g^{-1}(y_{dj}))+\sum_{j\in r_{d}}E_{y_r}\big[h(g^{-1}(y_{dj}))\big|y_s\big]\Big\}.
\end{eqnarray*}
The conditional distribution (\ref{conditioned}) depends on the vector
$\psi=(\beta^{\prime},\theta^\prime)^{\prime}$
of unknown model parameters, which must be estimated, that is,
$$
E_{y_r}\big[h(g^{-1}(y_{dj}))\big|y_s\big]=E_{y_r}\big[h(g^{-1}(y_{dj}))\big|y_s;\psi\big].
$$
Let
$\hat\psi=(\hat\beta^{\prime},\hat\theta^\prime)^{\prime}$ be an estimator based on sample data $y_s$.
The EBP of $\delta_d$ is
$$
\hat\delta_d^{eb}=\frac{1}{N_d}\Big\{\sum_{j\in s_{d}}h(g^{-1}(y_{dj}))
+\sum_{j\in r_{d}}E_{y_r}\big[h(g^{-1}(y_{dj}))\big|y_s;\hat\psi\big]\Big\}.
$$
For a general function $h$, the expected value above might be not tractable analytically.
When this occurs, the following Monte Carlo procedure can be applied.
\begin{itemize}
\item[(a)]
Estimate the unknown parameter $\psi=(\beta^{\prime},\theta^\prime)^{\prime}$ using sample data $(y_s,X_s)$.
\item[(b)]
Replacing $\psi=(\beta^{\prime},\psi^\prime)^\prime$ by the
estimate $\hat\psi=(\hat\beta^\prime,\hat\theta^\prime)^\prime$
obtained in (a), draw $L$ copies of each non-sample variable
$y_{dj}$ as
$$
y_{dj}^{(\ell)}\sim N_2(\hat\mu_{dj|s},\hat{V}_{d|s}),\quad j\in r_{d},\,\, d=1,\ldots,D,\,\, \ell=1,\ldots,L.
$$
where
$$
\hat\mu_{dj|s}=\left\{\begin{array}{ll}
X_{dj}\hat\beta+\hat{V}_{ud}Z_{ds}^\prime \Big\{\hat{V}_{eds}^{-1}-\hat{V}_{eds}^{-1}Z_{ds}\Big(\hat{V}_{ud}^{-1}+n_d\hat{V}_{edj}^{-1}\Big)^{-1}Z_{ds}^\prime \hat{V}_{eds}^{-1}\Big\}(y_{ds}-X_{ds}\hat\beta)&
\mbox{if }\, n_d\neq0,
\\
X_{dj}\hat\beta& \mbox{if }\, n_d=0,
\end{array}\right.
$$
and
$$
\hat{V}_{d|s}=\left\{\begin{array}{ll}
\hat{V}_{ud}+\hat{V}_{edj}-n_d\hat{V}_{ud}\hat{V}_{edj}^{-1}\hat{V}_{ud}
+n_d^2\hat{V}_{ud}\hat{V}_{edj}^{-1}\Big(\hat{V}_{ud}^{-1}+n_d\hat{V}_{edj}^{-1}\Big)^{-1}\hat{V}_{edj}^{-1}\hat{V}_{ud}&
\mbox{if }\, n_d\neq0,
\\
\hat{V}_{ud}+\hat{V}_{edj}& \mbox{if }\, n_d=0.
\end{array}\right.
$$
\item[(c)]
The Monte Carlo approximation of the expected value is
$$
E_{y_r}\big[h(g^{-1}(y_{dj}))|y_s;\hat\psi\big]\approx\frac{1}{L}\sum_{\ell=1}^L
h\big(g^{-1}(y_{dj}^{(\ell)})\big),\,\,\, j\in r_{d},\,d=1,\ldots,D.
$$
The Monte Carlo approximation of the EBP of $\delta_d$ is
\begin{equation}\label{deltad}
\hat\delta_d^{eb}\approx
\frac{1}{L}\sum_{\ell=1}^L\delta_d^{(\ell)},\,\,\,
\delta_d^{(\ell)}=\frac{1}{N_d}\Big(\sum_{j\in s_{d}}h(g^{-1}(y_{dj}))+\sum_{j\in r_{d}} h\big(g^{-1}(y_{dj}^{(\ell)})\big)\Big).
\end{equation}
\end{itemize}

\noindent{\bf Remark \ref{sect.ebp2NERaditive}.1.}
In many practical cases the values of the auxiliary variables are not available for all the population units.
If in addition some of the variables are continuous, the EBP method is not applicable.
An important particular case, where this method is applicable, is when the number of values of the vector of auxiliary variables is finite.
More concretely, suppose that the covariates are categorical such that $X_{dj}\in\{X_{01},\ldots,X_{0T}\}$, then we can calculate $\delta_d^{(\ell)}$ as
\begin{equation}\label{deltalfind}
\delta_d^{(\ell)}=\frac{1}{N_d}\left[\sum_{j=1}^{n_{d}}h(g^{-1}(y_{dj}))
+
\sum_{t=1}^{T}\sum_{j=1}^{N_{dt}-n_{dt}} h\big(g^{-1}(y_{dj}^{(\ell)})\big)\right],
\end{equation}
where $N_{dt}=\#\{j\in U_{d}:\,X_{dj}=X_{0t}\}$ is available from external data sources (aggregated auxiliary information), $n_{dt}=\#\{j\in s_{d}:\,X_{dj}=X_{0t}\}$, $y_{dtj}^{(\ell)}\sim N_2(\hat\mu_{dt|s},\hat{V}_{d|s})$, $d=1,\ldots,D$, $j=1,\ldots,N_{dt}-n_{dt}$, $t=1,\ldots,T$,  $\ell=1,\ldots,L$, where
$$
\hat\mu_{dj|s}=\left\{\begin{array}{ll}
X_{0t}\hat\beta+\hat{V}_{ud}Z_{ds}^\prime \Big\{\hat{V}_{eds}^{-1}-\hat{V}_{eds}^{-1}Z_{ds}\Big(\hat{V}_{ud}^{-1}+n_d\hat{V}_{edj}^{-1}\Big)^{-1}Z_{ds}^\prime \hat{V}_{eds}^{-1}\Big\}(y_{ds}-X_{ds}\hat\beta)&
\mbox{if }\, n_d\neq0,
\\
X_{0t}\hat\beta& \mbox{if }\, n_d=0,
\end{array}\right.
$$
and $\hat{V}_{d|s}$ was defined above.
%
%
%
%
%
%
%
\subsection{EBPs of non additive parameters}\label{sect.ebp2NER}
Let $z_{dj}=(z_{dj1},z_{dj2})^{\prime}$ be a vector of continuous positive variables measured on the sample unit $j$ of domain $d$, $d=1,\ldots,D$, $j=1,\ldots,n_d$.
Define $z_d=\underset{1\leq j \leq n_d}{\hbox{col}}(z_{dj})$.
This section consider domain $2\times1$ or $1\times 1$ parameters that can be written in the form
\begin{equation}\label{domparam}
\delta_d=h(z_{d}),
\end{equation}
where $h$ is a known measurable function $R^{2N_d}\mapsto R^t$, $t=1,2$. A domain parameter is the ratio
\begin{equation}\label{ratio}
R_{d}=\frac{\overline{Z}_{d1}}{\overline{Z}_{d1}+\overline{Z}_{d2}},\quad\mbox{where }\,
h(z_{d})=\frac{\sum_{j=1}^{N_d}z_{dj1}}{\sum_{j=1}^{N_d}(z_{dj1}+z_{dj2})}.
\end{equation}
As in Section \ref{sect.ebp2NERaditive}, we assume that there exist a one-to-one transformation $g:R^2\mapsto R^2$ such that $y_{dj}=g(z_{dj})$ follows the BNER model (\ref{modpop2NERdj}).
We further assume that $g$ is separable, i.e.
$$
y_{dj}=g(z_{dj})=\big(g_1(z_{dj1}),g_2(z_{dj2})\big)^\prime,\quad
z_{dj}=g^{-1}(y_{dj})=\big(g_1^{-1}(y_{dj1}),g_2^{-1}(y_{dj2})\big)^\prime,
$$
where $g_1:(0,\infty)\mapsto R$ and $g_2:(0,\infty)\mapsto R$ are one-to-one functions.
For easy of notation, we define the $2n_d\times 1$ vectors
$$
z_d=g^{-1}(y_d)=\underset{1\leq j \leq n_d}{\hbox{col}}\big(g^{-1}(y_{dj})\big),\quad d=1,\ldots,D.
$$
We can write (\ref{domparam}) and (\ref{ratio}) as functions of $y_{dj1}$ and $y_{dj2}$, i.e.
$$
\delta_d=h(g^{-1}(y_{d})),\quad
R_d=\frac{\sum_{j=1}^{N_d}g^{-1}_1(y_{dj1})}{\sum_{j=1}^{N_d}\big(g^{-1}_1(y_{dj1})+g^{-1}_2(y_{dj2})\big)}.
$$
The BP of $\delta_d$ is
$$
\hat\delta_d^B=E_{y_r}\big[h(g^{-1}(y_{d}))|y_s\big].
$$
The conditional distribution (\ref{conditioned}) depends on the vector
$\psi=(\beta^{\prime},\theta^\prime)^{\prime}$
of unknown model parameters, which must be estimated, that is,
$$
E_{y_r}\big[h(g^{-1}(y_{d}))|y_s\big]=E_{y_r}\big[h(g^{-1}(y_{d}))|y_s;\psi\big].
$$
Let
$\hat\psi=(\hat\beta^{\prime},\hat\theta^\prime)^{\prime}$ be an estimator based on sample data $y_s$.
The EBP of $\delta_d$ is
$$
\hat\delta_d^{eb}=E_{y_r}\big[h(g^{-1}(y_{d}))|y_s;\hat\psi\big].
$$
For a general function $h$, the expected value above might be not tractable analytically.
When this occurs, the following Monte Carlo procedure can be applied.
\begin{itemize}
\item[(a)]
Estimate the unknown parameter $\psi=(\beta^{\prime},\theta^\prime)^{\prime}$ using sample data $y_s$.
\item[(b)]
Replacing $\psi=(\beta^{\prime},\psi^\prime)^\prime$ by the
estimate $\hat\psi=(\hat\beta^\prime,\hat\theta^\prime)^\prime$
obtained in (a), draw $L$ copies of each non-sample variable
$y_{dj}$ as
$$
y_{dj}^{(\ell)}\sim N_2(\hat\mu_{dj|s},\hat{V}_{d|s}),\quad j\in r_{d},\,\, d=1,\ldots,D,\,\, \ell=1,\ldots,L.
$$
where
$$
\hat\mu_{dj|s}=\left\{\begin{array}{ll}
X_{dj}\hat\beta+\hat{V}_{ud}Z_{ds}^\prime \Big\{\hat{V}_{eds}^{-1}-\hat{V}_{eds}^{-1}Z_{ds}\Big(\hat{V}_{ud}^{-1}+n_d\hat{V}_{edj}^{-1}\Big)^{-1}Z_{ds}^\prime \hat{V}_{eds}^{-1}\Big\}(y_{ds}-X_{ds}\hat\beta)&
\mbox{if }\, n_d\neq0,
\\
X_{dj}\hat\beta& \mbox{if }\, n_d=0,
\end{array}\right.
$$
$$
\hat{V}_{d|s}=\left\{\begin{array}{ll}
\hat{V}_{ud}+\hat{V}_{edj}-n_d\hat{V}_{ud}\hat{V}_{edj}^{-1}\hat{V}_{ud}
+n_d^2\hat{V}_{ud}\hat{V}_{edj}^{-1}\Big(\hat{V}_{ud}^{-1}+n_d\hat{V}_{edj}^{-1}\Big)^{-1}\hat{V}_{edj}^{-1}\hat{V}_{ud}&
\mbox{if }\, n_d\neq0,
\\
\hat{V}_{ud}+\hat{V}_{edj}& \mbox{if }\, n_d=0.
\end{array}\right.
$$
\item[(c)]
Construct the vectors
$$
y_{dr}^{(\ell)}=\underset{j\in r_d}{\hbox{col}}\big(y_{dj}^{(\ell)}\big),\quad
y_{ds}^{(\ell)}=\underset{j\in s_d}{\hbox{col}}\big(y_{dj}\big),\quad
y_{d}^{(\ell)}=\big(y_{ds}^{(\ell)\prime},y_{dr}^{(\ell)\prime}\big)^\prime.
$$
\item[(d)]
The Monte Carlo approximation of the EBP of $\delta_d$ is
$$
\hat\delta_d^{eb}\approx\frac{1}{L}\sum_{\ell=1}^Lh(g^{-1}(y_{d}^{(\ell)})),\quad
d=1,\ldots,D.
$$
\end{itemize}
\noindent{\bf Remark \ref{sect.ebp2NER}.1.}
If the values of the auxiliary variables are not available for all the population units and some of them are continuous, the EBP method is not applicable.
This situation can be overcome when the covariates are categorical such that $X_{dj}\in\{X_{01},\ldots,X_{0T}\}$.
Then, we can write the elements of $y$ as
$y_{dtj}=(y_{dtj1},y_{dtj2})^\prime$, where $d$, $i$ and $j$ denote domain, category and individual respectively.
We can approximate $\hat{R}_d^{eb}$ as
\begin{equation}\label{Rdebp}
\hat{R}_d^{eb}=\frac{1}{L}\sum_{\ell=1}^{L}\frac{
\sum_{j=1}^{n_{d}}g_1^{-1}(y_{dj1})+\sum_{t=1}^{T}\sum_{j=1}^{N_{dt}-n_{dt}} g^{-1}_1(y_{dtj1}^{(\ell)})
}{
\sum_{j=1}^{n_{d}}(g^{-1}_1(y_{dj1})+g^{-1}_2(y_{dj2}))+\sum_{t=1}^{T}\sum_{j=1}^{N_{dt}-n_{dt}} (g^{-1}_1(y_{dtj1}^{(\ell)})+g^{-1}_2(y_{dtj2}^{(\ell)}))
},
\end{equation}
where $N_{dt}=\#\{j\in U_{d}:\,X_{dj}=X_{0t}\}$ is available from external data sources (aggregated auxiliary information), $n_{dt}=\#\{j\in s_{d}:\,X_{dj}=X_{0t}\}$, $y_{dtj}^{(\ell)}=(y_{dtj1}^{(\ell)},y_{dtj2}^{(\ell)})^\prime\sim N_2(\hat\mu_{dt|s},\hat{V}_{d|s})$, $d=1,\ldots,D$, $j=1,\ldots,N_{dt}-n_{dt}$, $t=1,\ldots,T$,  $\ell=1,\ldots,L$, where
$$
\hat\mu_{dt|s}=\left\{\begin{array}{ll}
X_{0t}\hat\beta+\hat{V}_{ud}Z_{ds}^\prime \Big\{\hat{V}_{eds}^{-1}-\hat{V}_{eds}^{-1}Z_{ds}\Big(\hat{V}_{ud}^{-1}+n_d\hat{V}_{edj}^{-1}\Big)^{-1}Z_{ds}^\prime \hat{V}_{eds}^{-1}\Big\}(y_{ds}-X_{ds}\hat\beta)&
\mbox{if }\, n_d\neq0,
\\
X_{0t}\hat\beta& \mbox{if }\, n_d=0,
\end{array}\right.
$$
and $\hat{V}_{d|s}$ was defined above.
%
%
%
%
%
\section{Parametric bootstrap MSE estimator}\label{sect.boot}
Analytical approximations to the MSE are difficult to derive in the case of complex parameters.
We therefore propose a parametric bootstrap MSE estimator by following the bootstrap method for finite populations of Gonz\'alez-Manteiga et al. (2007, 2008).
We present the case of additive domain parameters. The modifications to deal with non additive parameters are straightforward.
The steps for implementing this method are
\begin{itemize}
\item[1.]
Fit the model (\ref{modpop2NERdj}) to sample data $(y_s,X_s)$ and calculate an estimator
$\hat\phi=(\hat\beta^{\prime},\hat\theta^\prime)^{\prime}$ of $\psi=(\beta^{\prime},\theta^\prime)^{\prime}$.
\item[2.]
For $d=1,\ldots,D$, $j=1,\ldots,N_{d}$, generate independently $u_{d}^{*}\sim N(0,\hat{V}_{ud})$ and $e_{dj}^{*}\sim N(0,\hat{V}_{edj})$.
\item[3.]
Construct the bootstrap superpopulation model $\xi^*$ using $\{u_{d}^{*}\}$, $\{e_{dj}^{*}\}$,
$\{X_{dj}\}$ and $\hat{\beta}$, i.e
\begin{equation}\label{xi*}
\xi^{*}:\,
y_{dj}^{*}=X_{dj}\hat{\beta}+u_{d}^{*}+e_{dj}^{*},\,\,d=1,\ldots,D, j=1,\ldots,N_{d}.
\end{equation}
\item[4.]
Under the bootstrap superpopulation model (\ref{xi*}), generate a large number $B$ of i.i.d. bootstrap populations
$\{y_{dj}^{*(b)}:\,d=1,\ldots,D,  j=1,\ldots,N_{d}\}$ and calculate the bootstrap population parameters
$$
\delta_d^{*(b)}=\frac{1}{N_d}\sum_{j=1}^{N_{d}}h(g^{-1}(y_{dj}^{*(b)})),\quad b=1,\ldots,B.
$$
\item[5.]
From each bootstrap population $b$ generated in Step 4, take the sample with the same indices $s\subset U$ as the initial sample, and calculate the bootstrap EBPs, $\hat{\delta}_{d}^{eb*(b)}$, as described in
Section \ref{sect.ebp2NERaditive} using the bootstrap sample data $y_s^*$ and the known values $X_{dj}$.
\item[6.]
A Monte Carlo approximation to the theoretical bootstrap estimator
$$
MSE_*(\hat{\delta}_{d}^{eb*})=E_{\xi^*}\big[(\hat{\delta}_{d}^{eb*}-\delta_{d}^{*})(\hat{\delta}_{d}^{eb*}-\delta_{d}^{*})^\prime\big]
$$
is
\begin{equation}\label{mse*}
mse_*(\hat{\delta}_{d}^{eb*})=\frac{1}{B}\sum_{b=1}^B(\hat{\delta}_{d}^{eb*(b)}-\delta_{d}^{*(b)})(\hat{\delta}_{d}^{eb*(b)}-\delta_{d}^{*(b)})^\prime.
\end{equation}
The estimator (\ref{mse*}) is used to estimate $MSE(\hat{\delta}_d^{eb})$.
\end{itemize}

\section{Simulations}\label{sect.sim}
\subsection{Simulation 1}\label{sectsim1}
The target of Simulation 1 is to investigate the behavior of the EBPs based on the BNER model (\ref{modpop2NERdj}).
For this sake, we generate artificial population data as follows.
Take
$p_1=p_2=2$, $p=4$,
$\beta_1=(\beta_{11},\beta_{12})^\prime=(10,10)^\prime$,
$\beta_2=(\beta_{21},\beta_{22})^\prime=(10,10)^\prime$.
For $k=1,2$, $d=1,\ldots,D$,  $j=1,\ldots,n_d$, generate
$X_{dj}=\mbox{diag}(x_{dj1},x_{dj2})_{2\times 4}$, where
$x_{dj1}=(x_{dj11},x_{dj12})$, $x_{dj2}=(x_{dj21},x_{dj22})$,
$x_{dj11}=x_{dj21}=1$, $x_{dj12}\sim \mbox{Bin}(1,1/2)$, $x_{dj22}\sim \mbox{Bin}(1,1/2)$.
For $d=1,\ldots,D$, simulate ${u}_{d}\sim N_{2}(0,{V}_{ud})$ and
${e}_{dj}\sim N_{2}(0,{V}_{edj})$, where
$$
V_{ud}=\left(\begin{array}{cc}
\theta_1&\theta_3\sqrt{\theta_1}\sqrt{\theta_2}\\
\theta_3\sqrt{\theta_1}\sqrt{\theta_2}&\theta_2\\
\end{array}\right),\,\,\,
V_{ed}=\left(\begin{array}{cc}
\theta_4&\theta_{6}\sqrt{\theta_4}\sqrt{\theta_5}\\
\theta_{6}\sqrt{\theta_4}\sqrt{\theta_5}&\theta_5\\
\end{array}\right),
$$
with $\theta_1=0.75$, $\theta_2=1.00$, $\theta_4=0.50$, $\theta_5=0.75$ and $\theta_3=-0.8$, $\theta_6=0.8$.
Simulation 1 generates only 4 different matrices $X_{dj}$.
They are
$$
X_{dj}=\left(\begin{array}{cc|cc}
x_{dj11}&x_{dj12}&0&0\\
\hline 0&0&x_{dj21}&x_{dj22}\end{array}\right)\in\big\{
X_{01},X_{02},X_{03},X_{04}\big\} ,
$$
where
$$
X_{01}=\left(\begin{array}{cc|cc}
1&0&0&0\\
\hline 0&0&1&0\end{array}\right), X_{02}=\left(\begin{array}{cc|cc}
1&0&0&0\\
\hline 0&0&1&1\end{array}\right), X_{03}=\left(\begin{array}{cc|cc}
1&1&0&0\\
\hline 0&0&1&0\end{array}\right), X_{04}=\left(\begin{array}{cc|cc}
1&1&0&0\\
\hline 0&0&1&1\end{array}\right).
$$
For investigating the behavior of the predictors $\hat{A}^{eb}_d$ and $\hat{R}^{eb}_d$,
Simulation 1 takes the same transformations as in the application to real data, i.e.
$g_1(z_{dj1})=\log z_{dj1}$ and $g_2(z_{dj2})=\log z_{dj2}$, so that $z_{dj1}=\exp\{ y_{dj1}\}$ and $z_{dj2}=\exp\{y_{dj2}\}$.
Take $I=200$, $L=200$,  $N_d=200$ and $D=50$.

The steps of Simulation 1 are
\begin{enumerate}
\item[1.]
Generate $x_{djk}$, $d=1,\ldots,D$, $j=1,\ldots,N_d$, $k=1,2$.
Construct the population matrices $X_d$ and $Z_d$ of dimensions $2N_d\times p$ and $2N_d\times 2$ respectively.
For $d=1,\ldots,D$, $t=1,\ldots,T$, $T=4$, calculate
\begin{eqnarray*}
N_{dt}&=&\#\big\{j\in U_d:\, X_{j}=X_{0t}\big\}=\#\big\{j\in\iN:\,
j\leq N_d, X_{j}=X_{0t}\big\},
\\
n_{dt}&=&\#\big\{j\in s_d:\, X_{j}=X_{0t}\big\}=\#\big\{j\in\iN:\,
j\leq n_d, X_{j}=X_{0t}\big\}.
\end{eqnarray*}
\item[2.]
Repeat $I=200$ times ($i=1,\ldots,200$)
\begin{enumerate}
\item[2.1.]
Generate the populations random vectors  $u_d^{(i)}\sim N_2(0,V_{ud})$, $e_d^{(i)}\sim N_{2N_d}(0,V_{ed})$,
$y_{d}^{(i)}=X_{d}\beta+Z_du_{d}^{(i)}+e_{d}^{(i)}$, $d=1,\ldots,D$.
Calculate $z_{dj1}^{(i)}=\exp\{y_{dj1}^{(i)}\}$,  $z_{dj2}^{(i)}=\exp\{y_{dj2}^{(i)}\}$, $d=1,\ldots,D$, $j=1,\ldots,N_d$.
\item[2.2.]
Calculate the domain ratio parameters, i.e.
$$
A_d^{(i)}=\frac{1}{N_d}\sum_{j=1}^{N_d}\frac{z_{dj1}^{(i)}}{z_{dj1}^{(i)}+z_{dj2}^{(i)}},\quad
R_d^{(i)}=\frac{\sum_{j=1}^{N_d}z_{dj1}^{(i)} }{
\sum_{j=1}^{N_d}z_{dj1}^{(i)}+\sum_{j=1}^{N_d}z_{dj2}^{(i)}},\quad
d=1,\ldots,D.
$$
\item[2.3.]
Extract the sample $(y_{dj}^{(i)},X_{dj})$, $d=1,\ldots,D$,
$j=1,\ldots,n_d$, with $n_d=\{10, 25, 50, 100\}$.
\item[2.4.]
Calculate the REML estimators
$\hat{\beta}_{11}^{(i)},\hat{\beta}_{12}^{(i)},\hat{\beta}_{21}^{(i)},\hat{\beta}_{22}^{(i)},
\hat{\theta}_{1}^{(i)},\ldots,\hat{\theta}_{6}^{(i)}$.
\item[2.5]
For $d=1,\ldots,D$, $t=1,\ldots,T$, calculate
$$
\hat\mu_{dt|s}^{(i)}=
X_{0t}\hat\beta^{(i)}+\hat{V}_{ud}^{(i)}Z_{ds}^\prime
\Big\{\hat{V}_{eds}^{(i)-1}-\hat{V}_{eds}^{(i)-1}Z_{ds}\Big(\hat{V}_{ud}^{(i)-1}+n_d\hat{V}_{edj}^{(i)-1}\Big)^{-1}Z_{ds}^\prime
\hat{V}_{eds}^{(i)-1}\Big\}(y_{ds}-X_{ds}\hat\beta^{(i)}),
$$
$$
\hat{V}_{d|s}^{(i)}=
\hat{V}_{ud}^{(i)}+\hat{V}_{edj}^{(i)}-n_d\hat{V}_{ud}^{(i)}\hat{V}_{edj}^{(i)-1}\hat{V}_{ud}^{(i)}
+n_d^2\hat{V}_{ud}^{(i)}\hat{V}_{edj}^{(i)-1}\Big(\hat{V}_{ud}^{(i)-1}+n_d\hat{V}_{edj}^{(i)-1}\Big)^{-1}\hat{V}_{edj}^{(i)-1}\hat{V}_{ud}^{(i)}.
$$
\item[2.5.]
For $d=1,\ldots,D$, $j=1,\ldots,N_{dt}-n_{dt}$, $t=1,\ldots,T$,
$\ell=1,\ldots,L$, generate
$$
y_{dtj}^{(i\ell)}=(y_{dtj1}^{(i\ell)},y_{dtj2}^{(i\ell)})^\prime\sim
N_2(\hat\mu_{dt|s}^{(i)},\hat{V}_{d|s}^{(i)})
$$
and calculate $z_{dtj1}^{(i\ell)}=\exp\{y_{dtj1}^{(i\ell)}\}$, $z_{dtj2}^{(i\ell)}=\exp\{y_{dtj2}^{(i\ell)}\}$.
\item[2.6.]
For $d=1,\ldots,D$, calculate the EBPs $\hat{A}_d^{eb(i)}$ and
$\hat{R}_d^{eb(i)}$, i.e.
\begin{equation}\label{EBP1Sim2b}
\hat{A}_d^{eb(i)}=\frac{1}{LN_d}\sum_{\ell=1}^L\left[\sum_{j=1}^{n_{d}}\frac{z_{dj1}^{(i)}}{z_{dj1}^{(i)}+z_{dj2}^{(i)}}
+ \sum_{t=1}^{T}\sum_{j=1}^{N_{dt}-n_{dt}}
\frac{z_{dtj1}^{(i\ell)}}{z_{dtj1}^{(i\ell)}+z_{dtj2}^{(i\ell)}}\right],
\end{equation}
\begin{equation}\label{EBP2Sim2b}
\hat{R}_d^{eb(i)}=\frac{1}{L}\sum_{\ell=1}^{L}\frac{
\sum_{j=1}^{n_{d}}z_{dj1}^{(i)}+\sum_{t=1}^{T}\sum_{j=1}^{N_{dt}-n_{dt}}
z_{dtj1}^{(i\ell)} }{
\sum_{j=1}^{n_{d}}(z_{dj1}^{(i)}+z_{dj2}^{(i)})+\sum_{t=1}^{T}\sum_{j=1}^{N_{dt}-n_{dt}}
(z_{dtj1}^{(i\ell)}+z_{dtj2}^{(i\ell)}). },
\end{equation}
\end{enumerate}
\item[3.] For $\hat\eta_d^{(i)}\in\big\{\hat{A}_d^{eb(i)},\hat{R}_d^{eb(i)}\big\}$, $\eta^{(i)}_{d}\in\big\{A_d^{(i)},R_d^{(i)}\big\}$,
$d=1,\ldots,D$, calculate
$$
RE_{d}(\eta)=\Big(
\frac{1}{I}\sum_{i=1}^{I}(\hat{\eta}_{d}^{eb(i)}-\eta_{d}^{(i)})^2\Big)^{1/2},\,\,
B_{d}(\eta)=\frac{1}{I}\sum_{i=1}^{I}(\hat{\eta}_{d}^{eb(i)}-\eta_{d}^{(i)}),\,\,
\eta_{d}=\frac{1}{I}\sum_{i=1}^{I}\eta_{d}^{(i)},
$$
$$
RE(\eta)=\frac{1}{D}\sum_{d=1}^D RE_{d}(\eta),\quad
AB(\eta)=\sum_{d=1}^D|B_{d}(\eta)|,\quad
RRE_{d}(\eta)=\frac{RE_{d}(\eta)}{\eta_{d}}100,
$$
$$
RB_{d}(\eta)=\frac{B_{d}(\eta)}{\eta_{d}}100,\,\,
RRE=\frac{1}{D}\sum_{d=1}^D RRE_{d}(\eta),\,\,
RAB=\frac{1}{D}\sum_{d=1}^D|RB_{d}(\eta)|.
$$
\end{enumerate}
Tables \ref{sect.sim}.1 and \ref{sect.sim}.2 present the simulation results.
We observe that the EBPs are basically unbiased and that the MSEs decrease as the sample sizes $n_d$ increase.
These results indicate that the optimality properties of the BPs are inherited by the EBPs.

\renewcommand{\arraystretch}{1.2}
\begin{center}
\begin{tabular}{|c|c|rrrr|rrrr|}
\hline
$D$ & $\eta$ &$n_d=10$   &$n_d=25$   &$n_d=50$  &$n_d=100$   &   $n_d=10$    &$n_d=25$   &$n_d=50$  &$n_d=100$\\
\hline
\multirow{2}{*}{25}
& $\hat{A}^{eb}$ &    0.0007&     0.0007&   0.0004 &  0.0002&  0.0142&   0.0089&  0.0060& 0.0039   \\
& $\hat{R}^{eb}$ &    0.0022&     0.0014&   0.0010 &  0.0010&  0.0341&   0.0244&  0.0188& 0.0139   \\
\hline
\multirow{2}{*}{50}
& $\hat{A}^{eb}$ &   0.0007&     0.0005&    0.0004&   0.0002&   0.0143&     0.0090&     0.0062&   0.0039\\
& $\hat{R}^{eb}$ &   0.0018&     0.0014&    0.0011&   0.0007&   0.0336&     0.0242&     0.0190&   0.0142\\
\hline
\multirow{2}{*}{100}
& $\hat{A}^{eb}$ &   0.0008&  0.0005&    0.0003&   0.0002 &   0.0141& 0.0089&     0.0062&  0.0039\\
& $\hat{R}^{eb}$ &   0.0023&  0.0014&    0.0011&    0.0008&   0.0333& 0.0235&     0.0191&  0.0140 \\
\hline
\multirow{2}{*}{200}
& $\hat{A}^{eb}$ &   0.0008& 0.0005&   0.0004 &   0.0002& 0.0141&  0.0090  & 0.0062&  0.0039 \\
& $\hat{R}^{eb}$ &   0.0021& 0.0014&  0.0011  &   0.0008& 0.0328&  0.0240  & 0.0190&  0.0141 \\
\hline
\multicolumn{10}{c}{Table \ref{sect.sim}.1. $AB(\eta)$ (left) and $RE(\eta)$ (right), with $D=50$, $N_d=200$.}
\end{tabular}
\end{center}

\renewcommand{\arraystretch}{1.2}
\begin{center}
\begin{tabular}{|c|c|rrrr|rrrr|}
\hline
$D$ & $\eta$ &$n_d=10$   &$n_d=25$   &$n_d=50$  &$n_d=100$   &   $n_d=10$    &$n_d=25$   &$n_d=50$  &$n_d=100$\\
\hline
\multirow{2}{*}{25}
& $\hat{A}^{eb}$ &  0.1383  &     0.1348&    0.0877&  0.0497&   2.8523&    1.7977&    1.2271&  0.8049  \\
& $\hat{R}^{eb}$ &  0.4534  &     0.2998&    0.2163&  0.2068&   7.0479&    5.1039&    3.9724&  2.9803   \\
\hline
\multirow{2}{*}{50}
& $\hat{A}^{eb}$ &    0.1475&   0.0944&    0.0809&   0.0420&  2.8817&     1.8205&     1.2523&   0.7920 \\
& $\hat{R}^{eb}$ &    0.3783&   0.2885&    0.2310&   0.1376&  7.0344&     5.0821&     3.9981&   2.9698\\
\hline
\multirow{2}{*}{100}
& $\hat{A}^{eb}$ &    0.1516&     0.1033&  0.0678&   0.0485&    2.8452&    1.7948&  1.2542   &   0.7899\\
& $\hat{R}^{eb}$ &    0.4773&     0.2867&    0.2330& 0.1658&    7.0110&    4.9162&   3.9954  &   2.9319\\
\hline
\multirow{2}{*}{200}
& $\hat{A}^{eb}$ &   0.1544 &  0.0933&  0.0762  &  0.0434  & 2.8288&  1.8007  &   1.2416  &   0.7883 \\
& $\hat{R}^{eb}$ &   0.4476 &  0.2824& 0.2223   &  0.1665  & 6.8930&  5.0264  &   3.9642  &   2.9435  \\
\hline
\multicolumn{10}{c}{Table \ref{sect.sim}.2. $RAB(\eta)$ (left) and $RRE(\eta)$ (right), in \%, with $D=50$, $N_d=200$.}
\end{tabular}
\end{center}

\subsection{Simulation 2}\label{sectsim2}
For investigating the behavior of the bootstrap-based MSE estimators of predictors $\hat{A}^{eb}_d$ and $\hat{R}^{eb}_d$,
Simulation 2 uses the same artificial population as Simulation 1.
It also applies the transformations
$g_1(z_{dj1})=\log z_{dj1}$ and $g_2(z_{dj2})=\log z_{dj2}$, so that $z_{dj1}=\exp\{ y_{dj1}\}$ and $z_{dj2}=\exp\{y_{dj2}\}$.
Take $I=200$, $L=200$,  $N_d=200$, $D=50$ and $n_d=10$.
The steps of Simulation 2 are
\begin{enumerate}
\item[1.]
For $D=50$, generate $x_{djk}$, $d=1,\ldots,D$, $j=1,\ldots,N_d$,
$k=1,2$. Construct the population matrices $X_{dj}$ of dimensions
$2\times p$.
\item[2.]
Take $MSE_{d}(\eta)=RE_{d}(\eta)^2$, $d=1,\ldots,D$, from the output
of Simulation 2.
\item[3.]
Repeat $I=200$ times ($i=1,\ldots,200$)
\begin{enumerate}
\item[3.1.]
Generate the populations random vectors  $u_d^{(i)}\sim N_2(0,V_{ud})$, $e_{dj}^{(i)}\sim N_{2}(0,V_{edj})$ and
$y_{dj}^{(i)}=X_{dj}\beta+u_{d}^{(i)}+e_{dj}^{(i)}$, $d=1,\ldots,D$
$j=1,\ldots,N_d$ ($N_d=200$).
Calculate $z_{dj1}^{(i)}=\exp\{y_{dj1}^{(i)}\}$,  $z_{dj2}^{(i)}=\exp\{y_{dj2}^{(i)}\}$, $d=1,\ldots,D$, $j=1,\ldots,N_d$.
\item[3.2.]
Extract the sample $(y_{dj},X_{dj})$, $d=1,\ldots,D$,
$j=1,\ldots,n_d$ ($n_d=10$).
\item[3.3.]
Calculate the REML estimators
$\hat{\beta}_{11}^{(i)},\hat{\beta}_{12}^{(i)},\hat{\beta}_{21}^{(i)},\hat{\beta}_{22}^{(i)},
\hat{\theta}_{1}^{(i)},\ldots,\hat{\theta}_{6}^{(i)}$.
\item[3.4.]
Repeat $B$ times, $B=\{ 50, 100, 200, 400\}$ ($b=1,\ldots,B$).
\begin{enumerate}
\item[(a)]
Generate the bootstrap population vectors $u_d^{*(ib)}\sim
N_2(0,\hat{V}_{ud})$, $e_{dj}^{*(ib)}\sim N_{2}(0,\hat{V}_{edj})$,
$$
y_{dj}^{*(ib)}=X_{dj}\hat\beta^{(i)}+u_{d}^{*(ib)}+e_{dj}^{*(ib)},\quad
d=1,\ldots,D,\,\, j=1,\ldots,N_d,
$$
where $\hat{V}_{ud}=V_{ud}(\hat\theta_1,\hat\theta_2,\hat\theta_3)$
and $\hat{V}_{edj}=V_{edj}(\hat\theta_4,\hat\theta_5,\hat\theta_6)$.
Calculate $z_{dj1}^{*(ib)}=\exp\{y_{dj1}^{*(ib)}\}$,  $z_{dj2}^{*(ib)}=\exp\{y_{dj2}^{*(ib)}\}$, $d=1,\ldots,D$, $j=1,\ldots,N_d$.
\item[(b)]
Calculate the bootstrap domain ratio parameters, i.e.
$$
A_d^{*(ib)}=\frac{1}{N_d}\sum_{j=1}^{N_d}\frac{z_{dj1}^{*(ib)}}{z_{dj1}^{*(ib)}+z_{dj2}^{*(ib)}},\quad
R_d^{*(ib)}=\frac{\sum_{j=1}^{N_d}z_{dj1}^{*(ib)} }{
\sum_{j=1}^{N_d} z_{dj1}^{*(ib)}+ \sum_{j=1}^{N_d}
z_{dj2}^{*(ib)}},\quad d=1,\ldots,D.
$$
\item[(c)]
Extract the bootstrap sample $(y_{dj}^{*(ib)},X_{dj})$,
$d=1,\ldots,D$, $j=1,\ldots,n_d$.
\item[(d)]
Calculate the bootstrap REML estimators
$\hat{\beta}_{11}^{*(ib)},\hat{\beta}_{12}^{*(ib)},\hat{\beta}_{21}^{*(ib)},\hat{\beta}_{22}^{*(ib)},
\hat{\theta}_{1}^{*(ib)},\ldots,\hat{\theta}_{6}^{*(ib)}$.
\item[(e)]
Calculate the EBPs $\hat{A}_d^{eb*(ib)}$, $\hat{R}_d^{eb*(ib)}$ and
$\hat{R}_d^{in*(ib)}$, $d=1,\ldots,D$, as in (\ref{EBP1Sim2b}) and
(\ref{EBP2Sim2b}), with $L=200$.
\end{enumerate}
\item[3.5]
For
$\hat\eta_{d}^{*(ib)}\in\big\{\hat{A}_d^{eb*(ib)},\hat{R}_d^{eb*(ib)}\big\}$,
$\eta^{*(ib)}_{d}\in\big\{A_d^{*(ib)},R_d^{*(ib)}\big\}$,
$d=1,\ldots,D$, calculate
$$
mse^{*(i)}_{d}=\frac{1}{B}\sum_{b=1}^{B}\big(\hat{\eta}_{d}^{eb*(ib)}-\eta_{d}^{*(ib)}\big)^2.
$$
\end{enumerate}
\item[4.] For $d=1,\ldots,D$, $\hat\eta\in\big\{\hat{A}_d^{eb},\hat{R}_d^{eb}\big\}$,  alculate
$$
RE_{d}(\hat\eta)=\Big(
\frac{1}{I}\sum_{i=1}^{I}(mse_{d}^{*(i)}-MSE_{d}(\hat\eta))^2\Big)^{1/2},\quad
B_{d}(\hat\eta)=\frac{1}{I}\sum_{i=1}^{I}(mse_{d}^{*(i)}-MSE_{d}(\hat\eta)),
$$
$$
RE(\hat\eta) =\frac{1}{D}\sum_{d=1}^D RE_{d}(\hat\eta),\quad AB(\hat\eta)
=\frac{1}{D}\sum_{d=1}^D|B_{d}(\hat\eta)|.
$$
$$
RRE_d(\hat\eta) =\frac{RE_{d}(\hat\eta)}{MSE_{d}(\hat\eta)}100,\quad RB_d(\hat\eta)
=\frac{B_{d}(\hat\eta)}{MSE_{d}(\hat\eta)}100.
$$
$$
RRE(\hat\eta) =\frac{1}{D}\sum_{d=1}^D RRE_{d}(\hat\eta),\quad RAB(\hat\eta)
=\frac{1}{D}\sum_{d=1}^D |RB_{d}(\hat\eta)| .
$$
\end{enumerate}
Tables \ref{sect.sim}.3 and \ref{sect.sim}.4 present the simulation results.
We observe that the MSEs of the MSE estimators decrease when the number of bootstrap resamples increases.
However the biases of the MSE estimators remain stable when $B$ increases.

\renewcommand{\arraystretch}{1.1}
\begin{center}
\begin{tabular}{|c|rrrrr|rrrrr|}
\hline $B$ &   50  &   100 &   200 &   300 &   400 &   50  &   100 &
200 &   300 &   400 \\  \hline
$\hat{A}^{eb}$  &   0.0224  &   0.0209  &    0.0212   &   0.0209&    0.0209   &   0.0592   &  0.0470  &    0.0396  &  0.0371  &   0.0349  \\
$\hat{R}^{eb}$  &  0.1055   &   0.0989   &    0.1026  &   0.1015&    0.1004   &   0.3307   &   0.2549  &    0.2086  &   0.1936  &   0.1818 \\
\hline \multicolumn{11}{c}{Table \ref{sect.sim}.3. $10^3
AB(\hat\eta)$ (left) and $10^3 RE(\hat\eta)$ (right) with $D=50$, $n_d=10$, $N_d=200$.}
\end{tabular}
\end{center}

\renewcommand{\arraystretch}{1.1}
\begin{center}
\begin{tabular}{|c|rrrrr|rrrrr|}
\hline $B$ &   50  &   100 &   200 &   300 &   400 &   50  &   100 &
200 &   300 &   400 \\  \hline
$\hat{A}^{eb}$  &   10.470   &    9.854   &   10.055   &  9.840  &    9.796   &   28.950  &   22.899  &   19.247  &   17.902   &   16.821 \\
$\hat{R}^{eb}$  &   9.515   &   8.968  &   9.312   &   9.163   &    9.087   &   29.673  &   22.915  &   18.788   &   17.416  &   16.368   \\
\hline \multicolumn{11}{c}{Table \ref{sect.sim}.4. $RAB(\hat\eta)$
(left) and $RRE(\hat\eta)$ (right), in \%, with $D=50$, $n_d=10$, $N_d=200$.}
\end{tabular}
\end{center}

\begin{figure}[!ht]
\begin{center}
\vspace*{-15mm}
\hspace*{-8mm}
\includegraphics[width=85mm,height=90mm,angle=0]{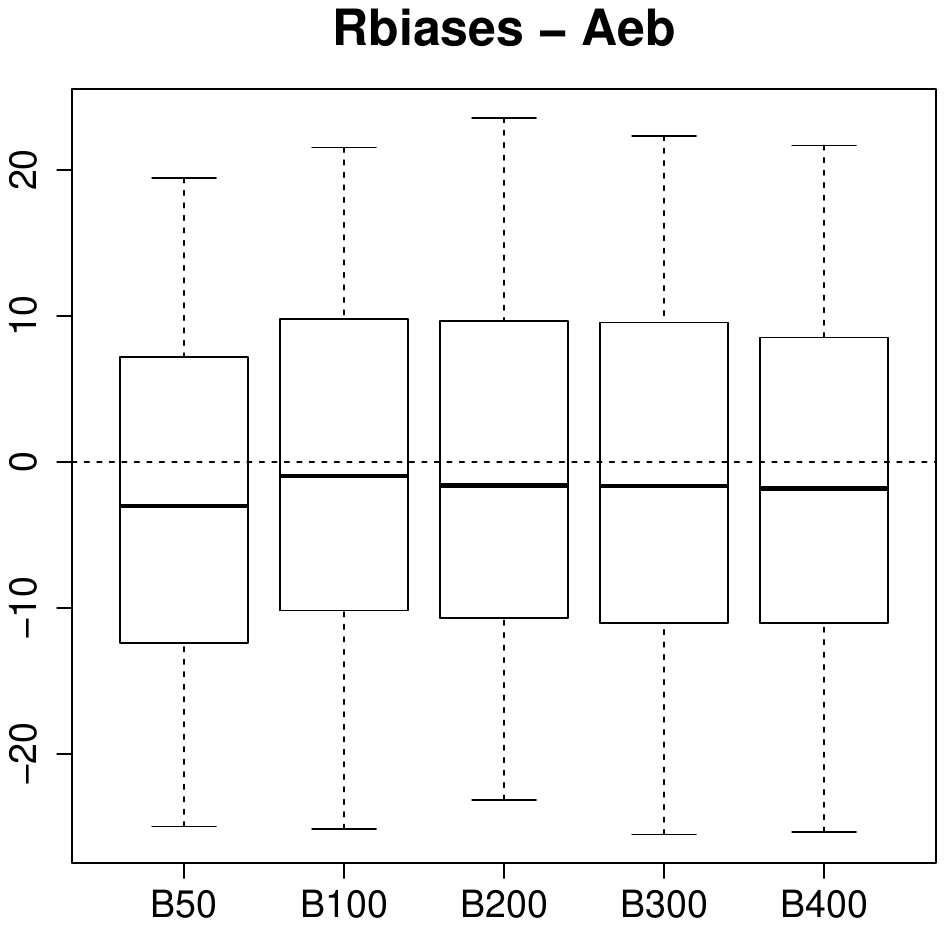}
\hspace*{-8mm}
\includegraphics[width=85mm,height=90mm,angle=0]{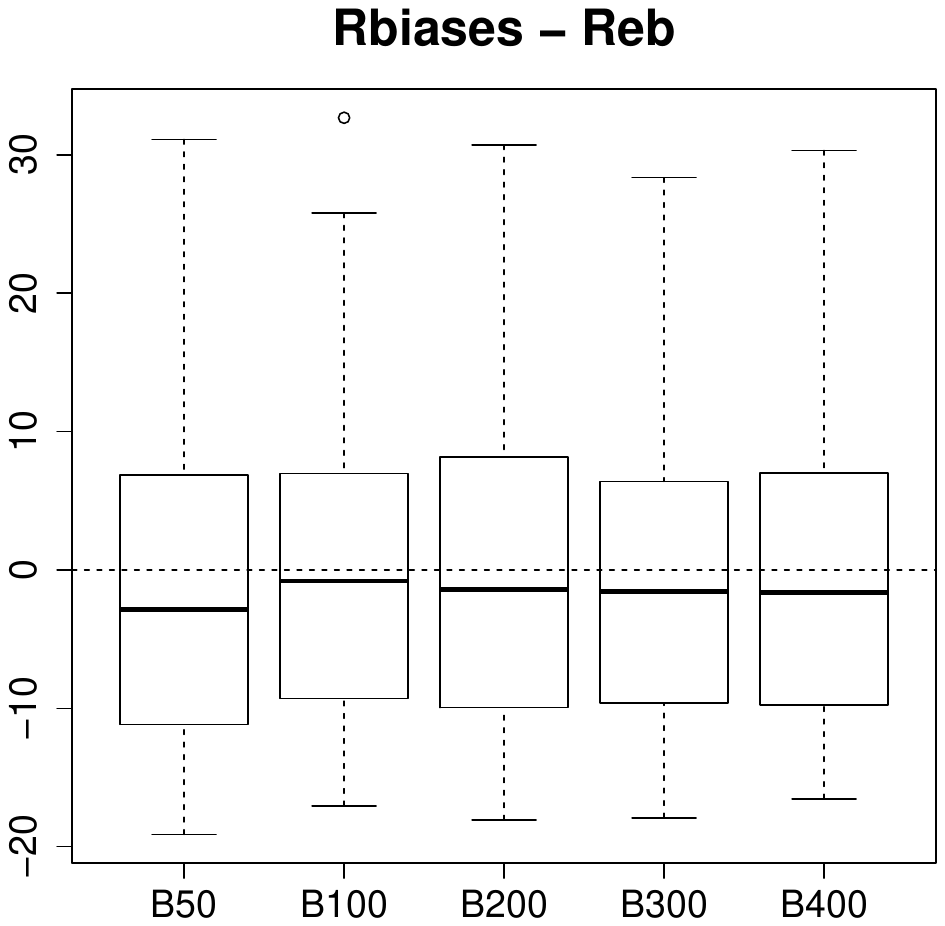}

\vspace*{-20mm}

\caption{Relative biases of the MSE estimators for $\hat{A}^{eb}_d$ (left) and $\hat{R}^{eb}_d$ (right).}
\label{boxplotbias}
\end{center}
\end{figure}

\begin{figure}[!ht]
\begin{center}
\vspace*{-15mm}
\hspace*{-8mm}
\includegraphics[width=85mm,height=90mm,angle=0]{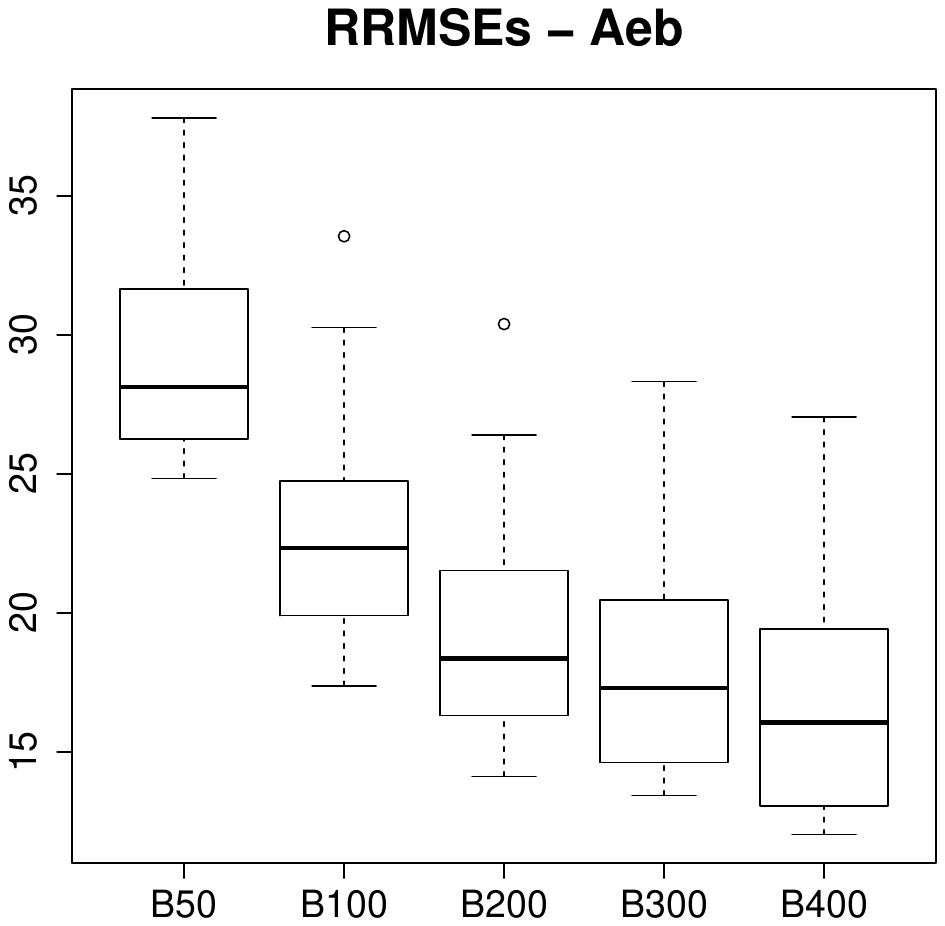}
\hspace*{-8mm}
\includegraphics[width=85mm,height=90mm,angle=0]{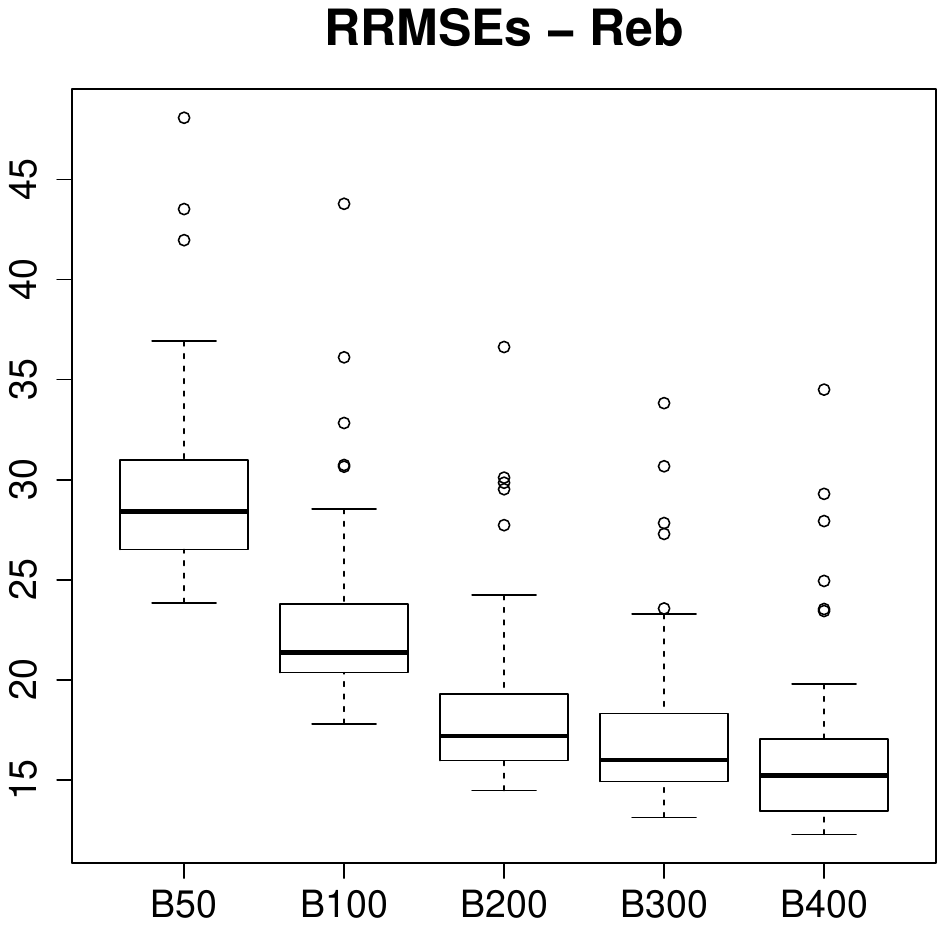}

\vspace*{-20mm}

\caption{Relative root-MSEs of the MSE estimators for $\hat{A}^{eb}_d$ (left) and $\hat{R}^{eb}_d$ (right).}
\label{boxplotRRMSE}
\end{center}
\end{figure}

Figures \ref{boxplotbias} contains the boxplots of the empirical relative biases, in \%,  of the parametric bootstrap estimators of the MSEs of the
predictors $\hat{A}^{eb}_d$ (left) and $\hat{R}^{eb}_d$ (right).
Figures \ref{boxplotRRMSE} presents the corresponding boxplots for the relative empirical relative root-MSEs.
The first figure shows that the bootstrap MSE estimators are rather unbiased, with a small tendency to under estimation.
The second figure suggests running around $B=400$ iterations in the bootstrap resampling procedure for obtaining good approximations to the MSEs of the EBPs.

%
%
%
%
%
%
%
\section{The Spanish Household Budget Survey}\label{sect.app}
The Spanish Household Budget Survey (SHBS) is annually carried out by the ``Instituto Nacional de Estad\'{\i}stica'' (INE), with the
objective of obtaining information on the nature and destination of the consumption expenses,
as well as on various characteristics related to the conditions of household life.
We deal with data from the SHBS of 2016.
This section presents an application of the new statistical methodology to the estimation of domain parameters defined as additive functions of two types of expense variables.

Let $z_{dj1}$ and $z_{dj2}$ be the food and non-food monthly expenses of household $j$ of domain (province) $d$.
The {\it domain means of food and non-food household monthly expenses} are
$$
\overline{Z}_{d1}=\frac{1}{N_d}\sum_{j=1}^{N_d}z_{dj1},\quad \overline{Z}_{d2}=\frac{1}{N_d}\sum_{j=1}^{N_d}z_{dj2},\quad d=1,\ldots,D,
$$
which are additive parameters with $h(z_{dj})=\frac{1}{N_d}(z_{dj1},z_{dj2})$.
The {\it domain ratios of food household monthly expense averages} are
$$
R_d =\frac{\overline{Z}_{d1}}{\overline{Z}_{d1}+\overline{Z}_{d2}},\quad d=1,\ldots,D,
$$
which are ratios of additive parameters.
The {\it domain averages of food household monthly expense ratios} are
$$
A_d = \frac{1}{N_d}\sum_{j=1}^{N_d}\frac{z_{dj1}}{z_{dj1}+z_{dj2}},\quad d=1,\ldots,D,
$$
which are a additive parameters with $h(z_{dj})=z_{dj1}/(z_{dj1}+z_{dj2})$.
As we do not have a Spanish census file dated around 2016, we estimate the domain parameters $\overline{Z}_{d1}$, $\overline{Z}_{d1}$, $A_d$ and $R_d$
by using the EBPs $\hat\delta_d^{eb}$ defined in (\ref{deltalfind}) and  (\ref{Rdebp}) respectively.
For this sake, we fit a BNER model to the target variables $z_{dj1}$ and $z_{dj2}$ with the following family composition auxiliary variables
\begin{itemize}\setlength\itemsep{-0.2em}
\item[FC1:]
Single person or adult couple with at least one members with age over 65,
\item[FC2:]
Other compositions with a single person or a couple without children,
\item[FC3:]
Couple with children under 16 years old or adult with children under 16 years old.
\item[FC4:]
Other households.
\end{itemize}
The variables FC are treated as a factor with reference category FC4.
We first fit a BNER model to the expenditure variables $z_{dj1}$ and $z_{dj2}$.
As the shape of the histogram estimators of the probability density functions of the model marginal residuals are slightly right-skewed, we apply the log transformation.
Therefore, we fit a BNER model to $y_{dj1}=\log z_{dj1}$ and $y_{dj2}=\log z_{dj2}$ with the same auxiliary variables.
For each target variable, $y_1$ and $y_2$, Table \ref{parameters} presents the estimates of the regression parameters and their standard errors.
It also presents the asymptotic $p$-values for testing the hypotheses $H_0:\beta_{kr}=0$, $k=1,2$, $r=1,2,3,4$.

\renewcommand{\arraystretch}{1}
\begin{table}[ht]
\centering
\begin{tabular}{|ll|rrrr|}
  \hline
$y$ & $x$-variable & estimate & $z$-value & st.error & $p$-value \\
  \hline
$y_1$ & intercept&  -0.80 & 45.29  & 0.02 & 0.00  \\
  &FC1 &   -0.36 & 28.94   & 0.01 & 0.00  \\
  &FC2 &  -0.61 & 49.23 & 0.01 & 0.00  \\
  &FC3 &-0.14 & 10.79 & 0.02 & 0.00 \\
  \hline
$y_2$ & intercept & 0.83 & 42.00 & 0.02 & 0.00  \\
&FC1 & -0.39 & 37.24& 0.01 & 0.00  \\
&FC2 & -0.29 & 27.93 & 0.01 & 0.00 \\
&FC3 & 0.01 & 1.14  & 0.01 & 0.25\\
   \hline
\end{tabular}
\caption {\label{parameters}Regression parameters and $p$-values.}
\end{table}

Table \ref{variance} presents the estimates of the variance and correlation parameters with their $95\%$ asymptotic confidence intervals.
This table shows that all the estimated parameters are significantly greater than zero.
We remark that correlations $\rho_u$ and $\rho_e$ are significantly greater than zero, so that the independent univariate modeling  of $y_1$ and $y_2$ is not appropriate.

\renewcommand{\arraystretch}{1}
\begin{table}[!ht]
\centering
\begin{tabular}{|l|rrr|}
  \hline
parameter & estimate& lower.lim & upper.lim \\
  \hline
$\sigma_{u1}^2$ & 0.013 & 0.007 & 0.018\\
$\sigma_{u2}^2$ & 0.018 & 0.010 & 0.025 \\
$\rho_u$ & 0.614 & 0.421 & 0.807 \\
$\sigma_{e1}^2$ &0.451 & 0.442 & 0.459  \\
$\sigma_{e2}^2$ & 0.318 & 0.312 & 0.324 \\
$\rho_e$ &  0.377 & 0.366 & 0.389  \\
   \hline
\end{tabular}
\caption{\label{variance}Variance and correlation parameters.}
\end{table}

\begin{figure}[!ht]
\begin{center}
\includegraphics[width=140mm,height=65mm]{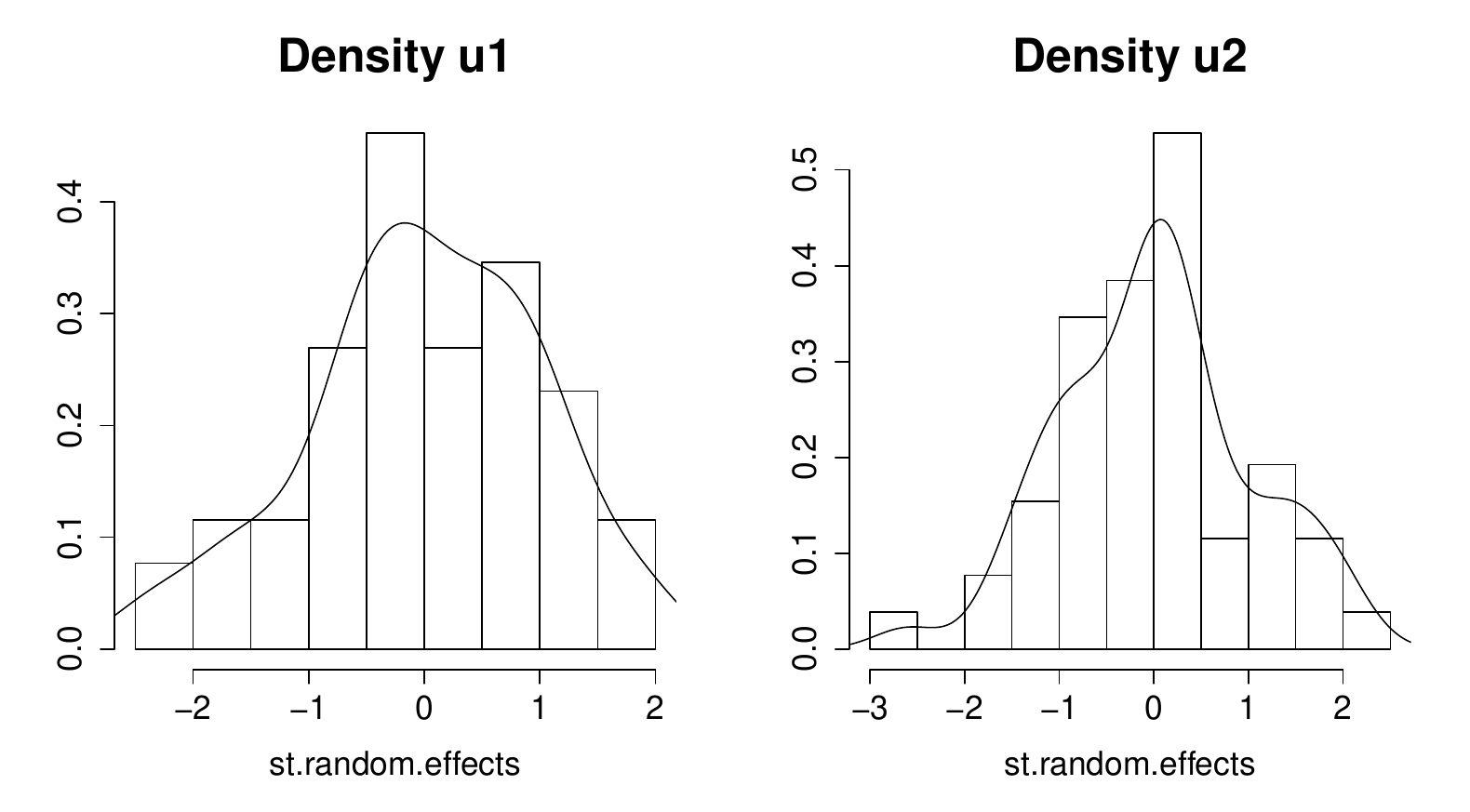}

\vspace*{-3mm}
\caption{Histograms of standardized random effects.}
\label{density_u_g}
\end{center}
\end{figure}

Figure \ref{density_u_g} plots the histograms of the $D=52$ standardized EBPs of the  random effects of the fitted BNER model for food (left) and non-food (right) expenditures.
The standardization of $\hat{u}_{d1}$ and $\hat{u}_{d2}$ is carried out by subtracting their mean value and dividing by their standard deviation.
It also prints the corresponding probability density function estimates.
The shapes of the densities are quite symmetrical, which indicates that the distributions of the random effects are not very far from the normal distributions.
Since $D$ is too small to obtain a good non-parametric estimate of the density functions, the definitive conclusions can not be drawn.

\begin{figure}[!ht]
\begin{center}
\includegraphics[width=140mm,height=65mm]{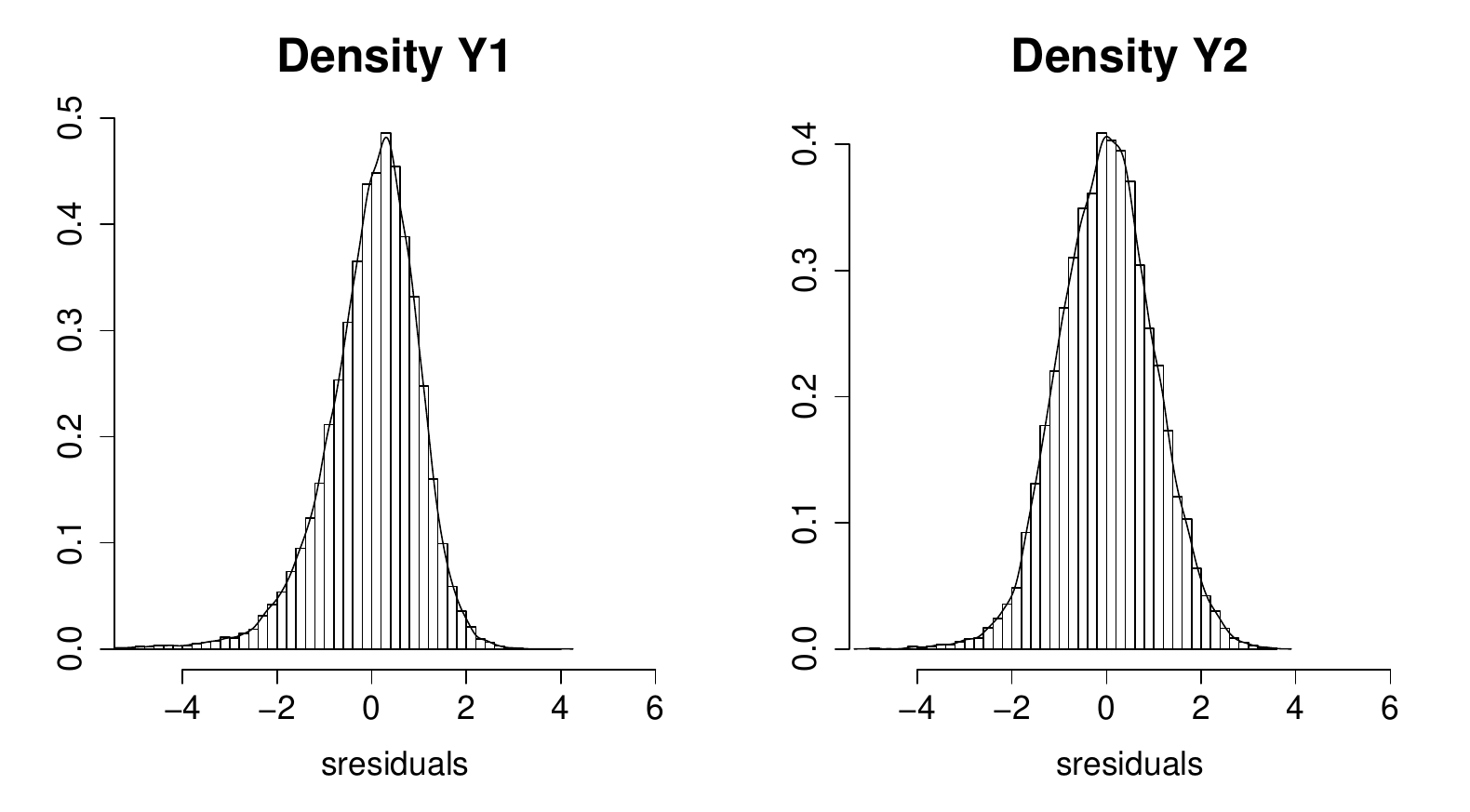}

\caption{Histograms of standardized residuals.}
\label{histogram_res_g}
\end{center}
\end{figure}

Figure \ref{histogram_res_g} plots the histograms of the $n=22010$ standardized residuals (sresiduals) of the fitted BNER model for the first (left) and second (right) response variables.
The standardization of $\hat{e}_{dj1}$ and $\hat{e}_{dj2}$ is carried out by subtracting their mean value and dividing by their standard deviation.
It also prints the corresponding probability density function estimates.
The curves of the estimated densities have longer left tails and slightly right-skewed shape.
Nevertheless, we could admit that the distributions of standardized residuals is not too far from normality.

\begin{figure}[!ht]
\begin{center}
\includegraphics[width=140mm,height=65mm]{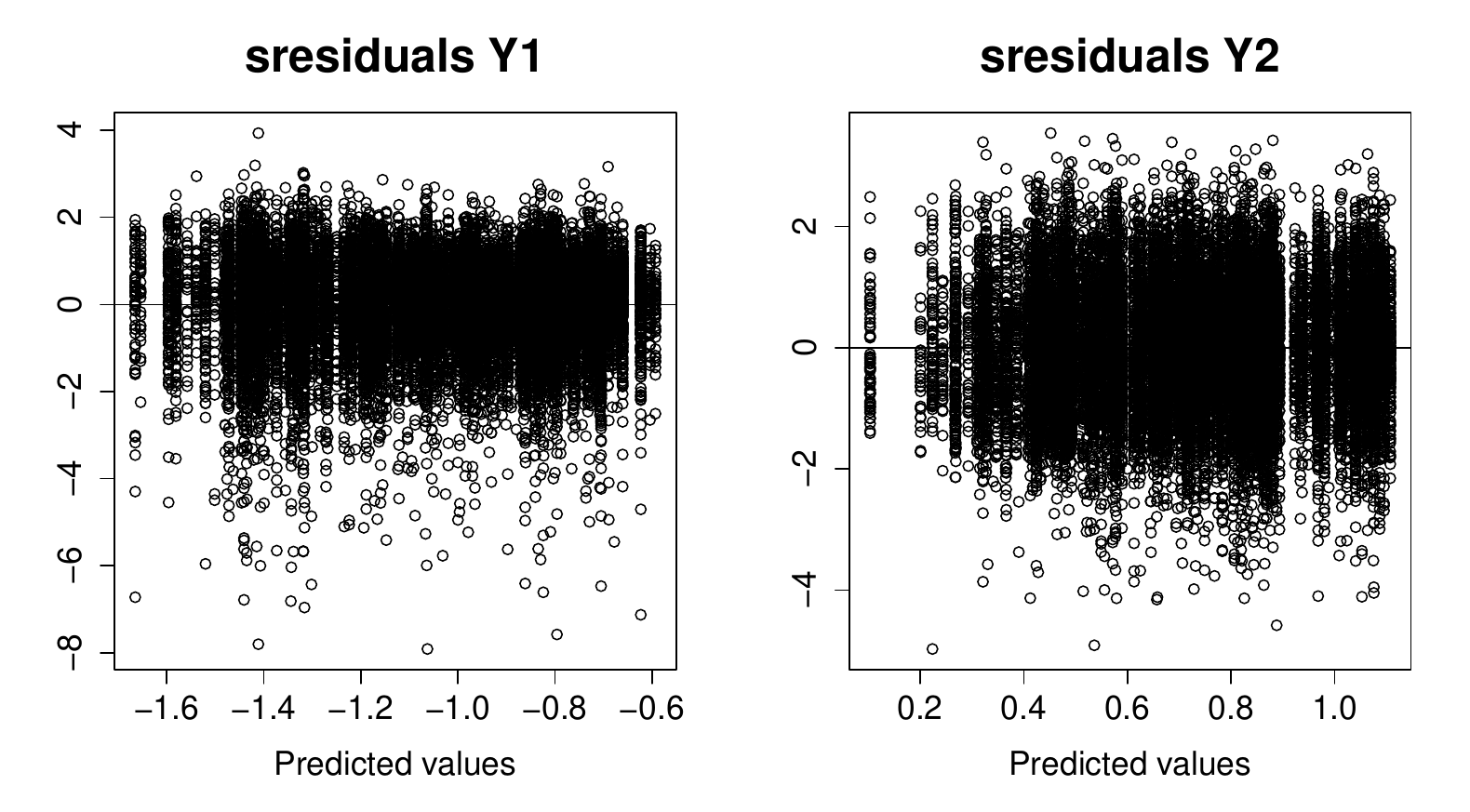}
\caption {\label{residuals_g1_g} Standardized residuals versus predicted values (in $10^3$ euros).}
\end{center}
\end{figure}

Figure \ref{residuals_g1_g} plots the standardized residuals versus the predicted values of the fitted BNER model,
which correspond to the logarithms of food (Y1) and non-food (Y2) expenditures.
In both cases, the main cloud of residuals is situated symmetrically around zero without any recognizable pattern.

\begin{figure}[!ht]
\begin{center}
\hspace*{-9mm}
\includegraphics[height=60mm,width=87mm,angle=0]{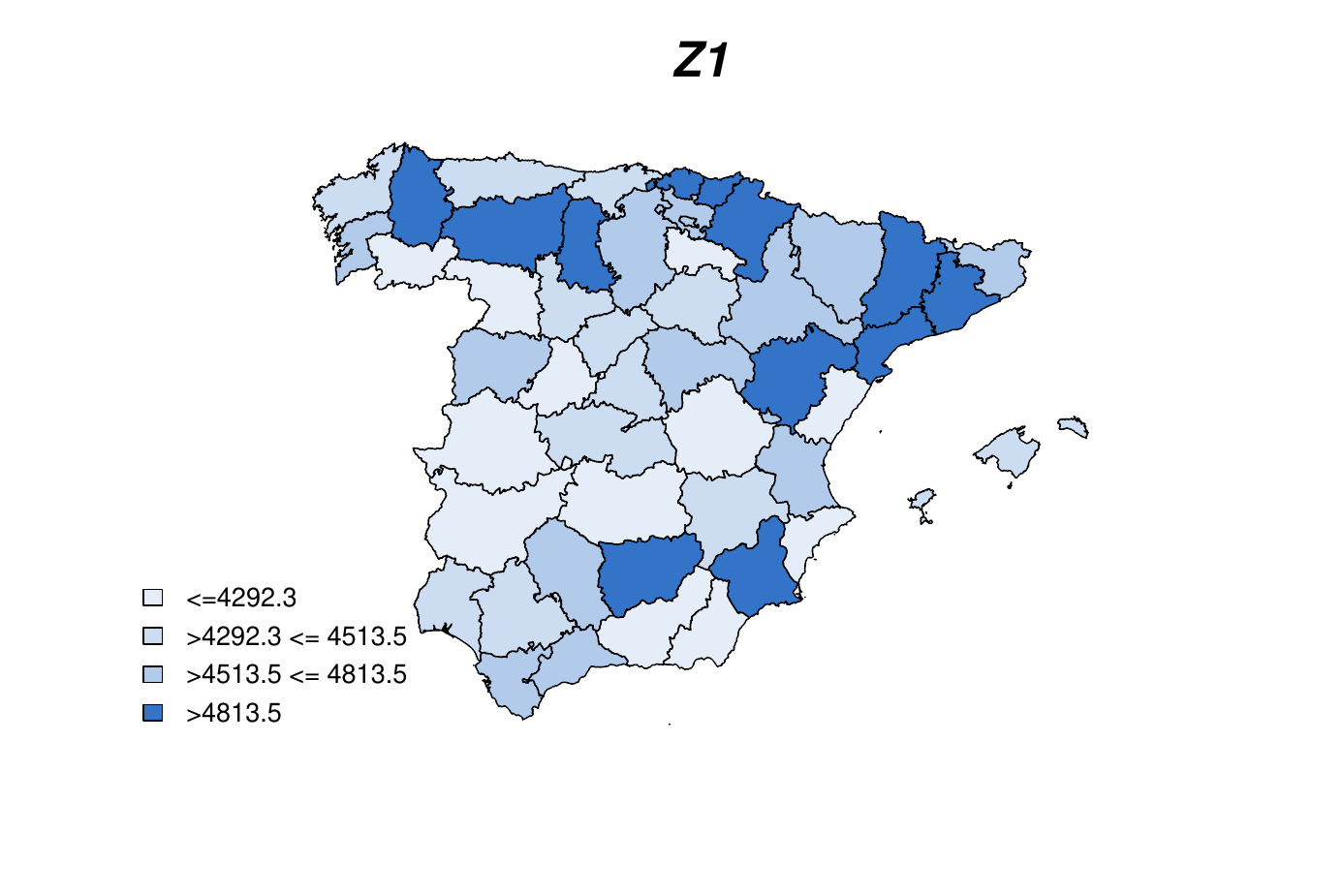}
\hspace*{-9mm}
\includegraphics[height=60mm,width=87mm,angle=0]{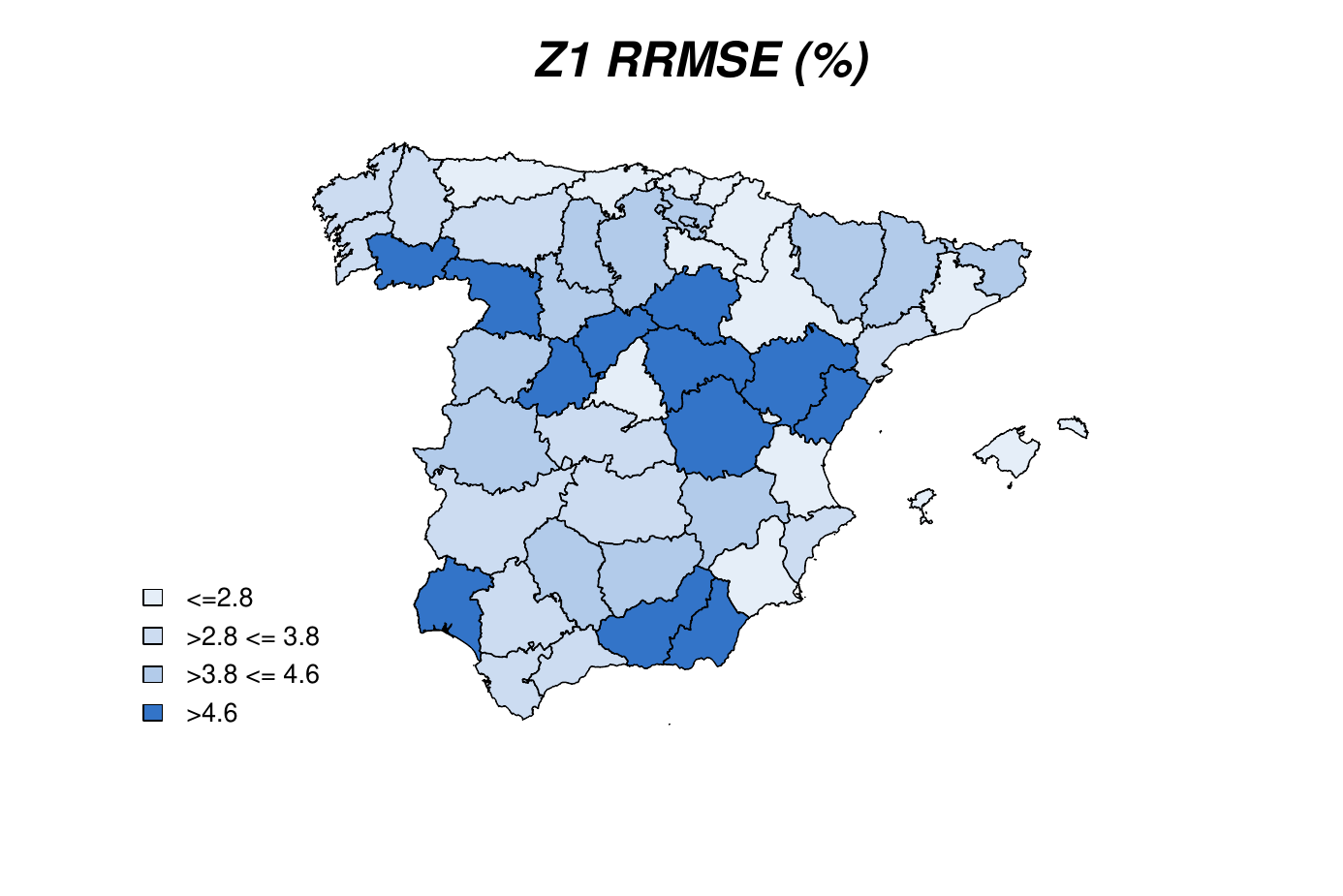}
\caption{Means $\hat{\overline{Z}}_{d1}^{eb}$ (left) and their relative root-MSEs in \% (right) of household annual expenditures in food by Spanish provinces.}
\label{map_meanFood_g}
\end{center}
\end{figure}

Figure \ref{map_meanFood_g} (left) maps the means of the household annual expenditures in food by Spanish provinces.
Figure \ref{map_meanFood_g} (right) maps the estimated relative root-MSEs (RRMSE) in \%.
These Figures show that expenditures on food is rather variable between provinces.

\begin{figure}[!ht]
\begin{center}
\hspace*{-9mm}
\includegraphics[height=60mm,width=87mm,angle=0]{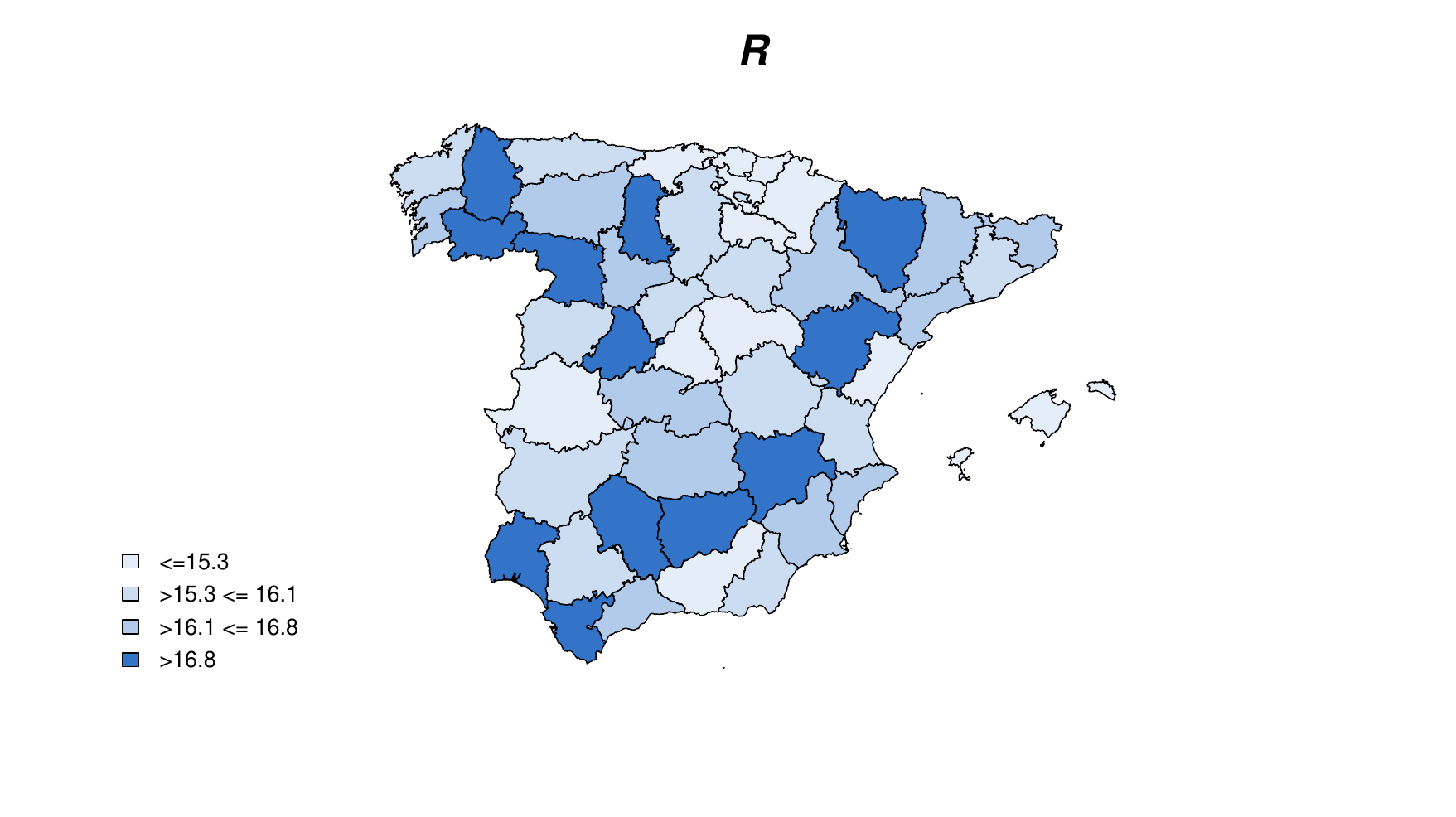}
\hspace*{-9mm}
\includegraphics[height=60mm,width=87mm,angle=0]{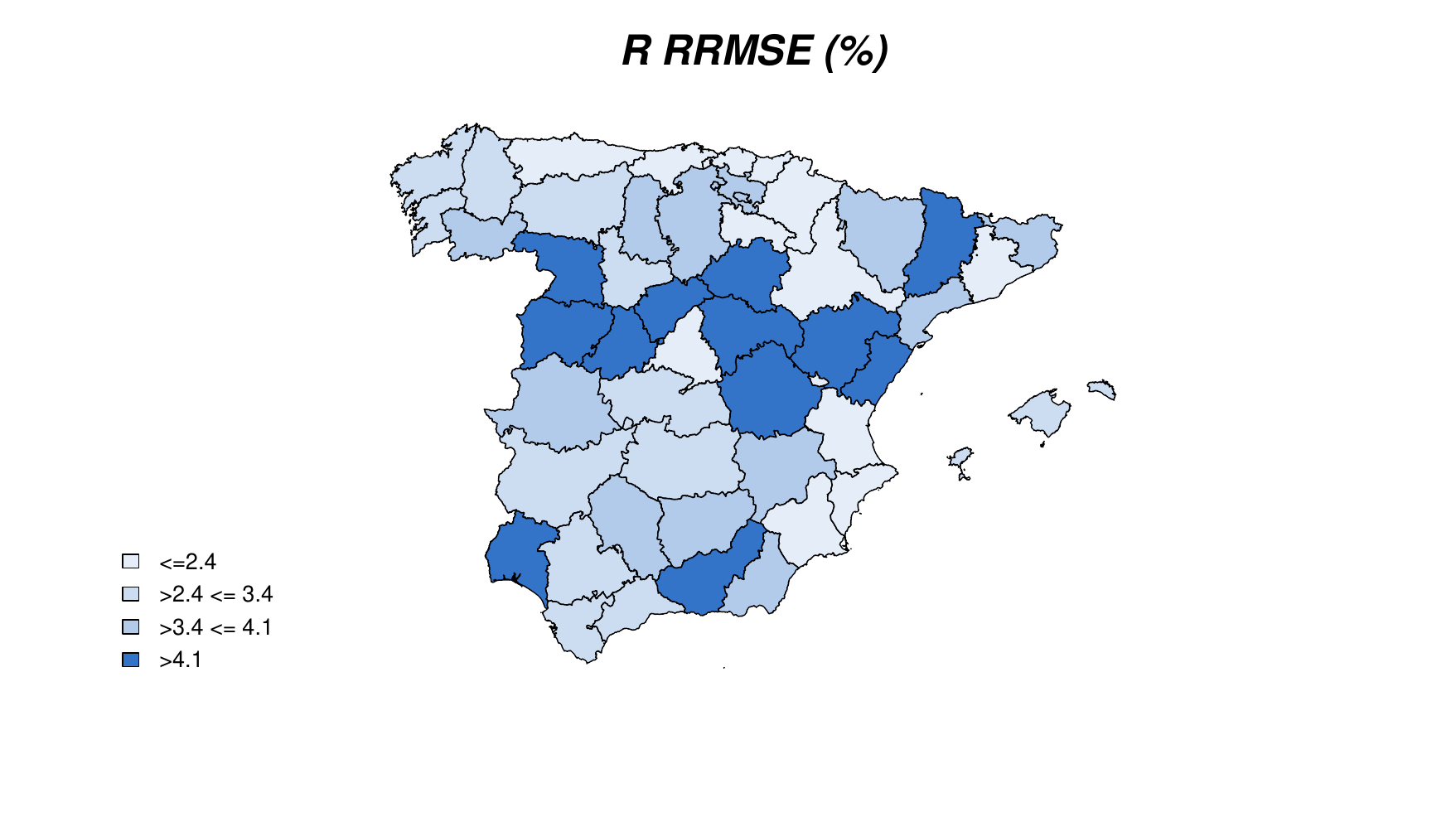}

\vspace*{-3mm}
\caption{Ratios $\hat{R}_{d}^{in}$ in \% of household annual expenditures in food by Spanish provinces.}
\label{map_percentage_g}
\end{center}
\end{figure}

\begin{figure}[!ht]
\begin{center}
\hspace*{-9mm}
\includegraphics[height=60mm,width=87mm,angle=0]{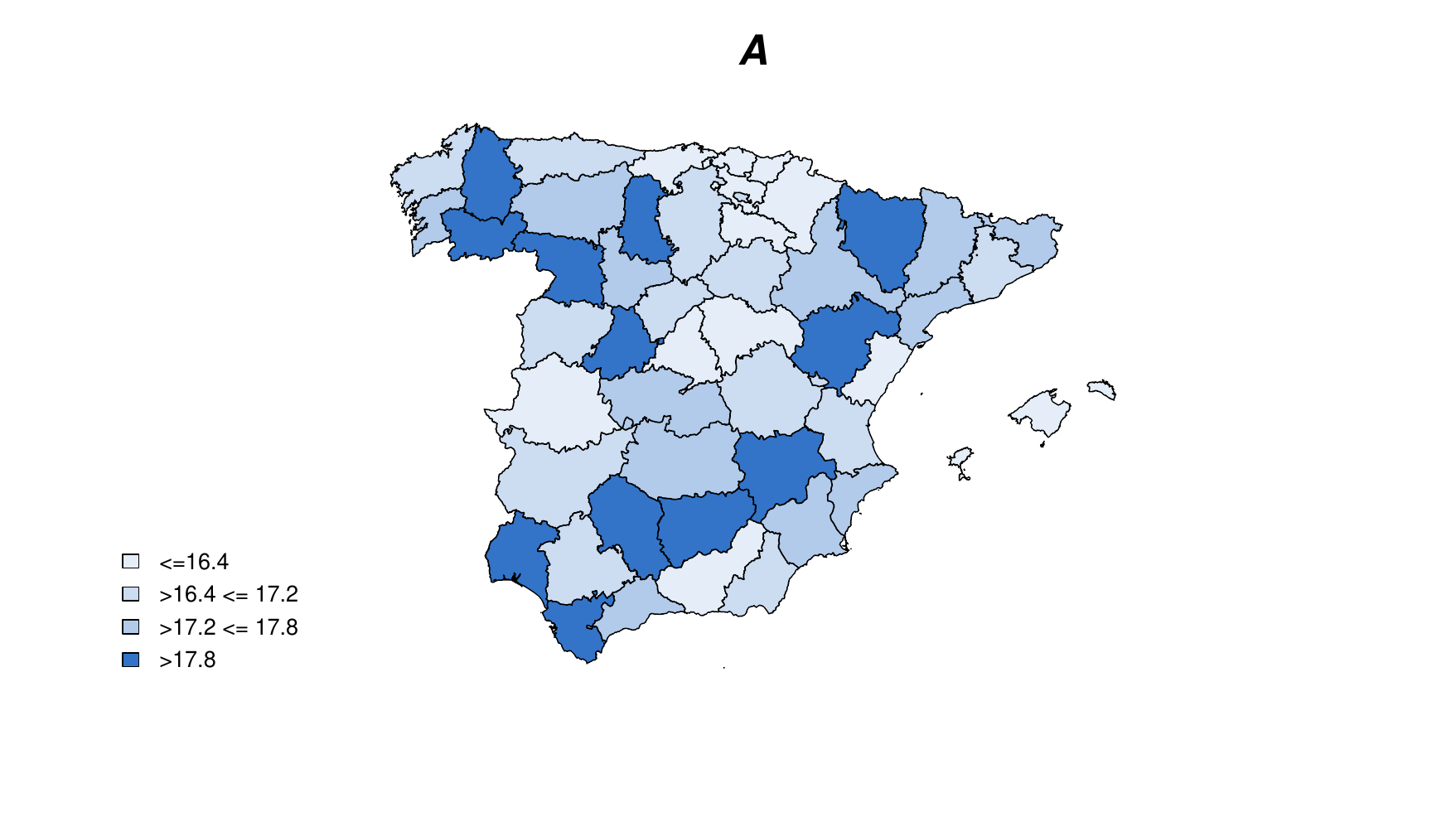}
\hspace*{-9mm}
\includegraphics[height=60mm,width=87mm,angle=0]{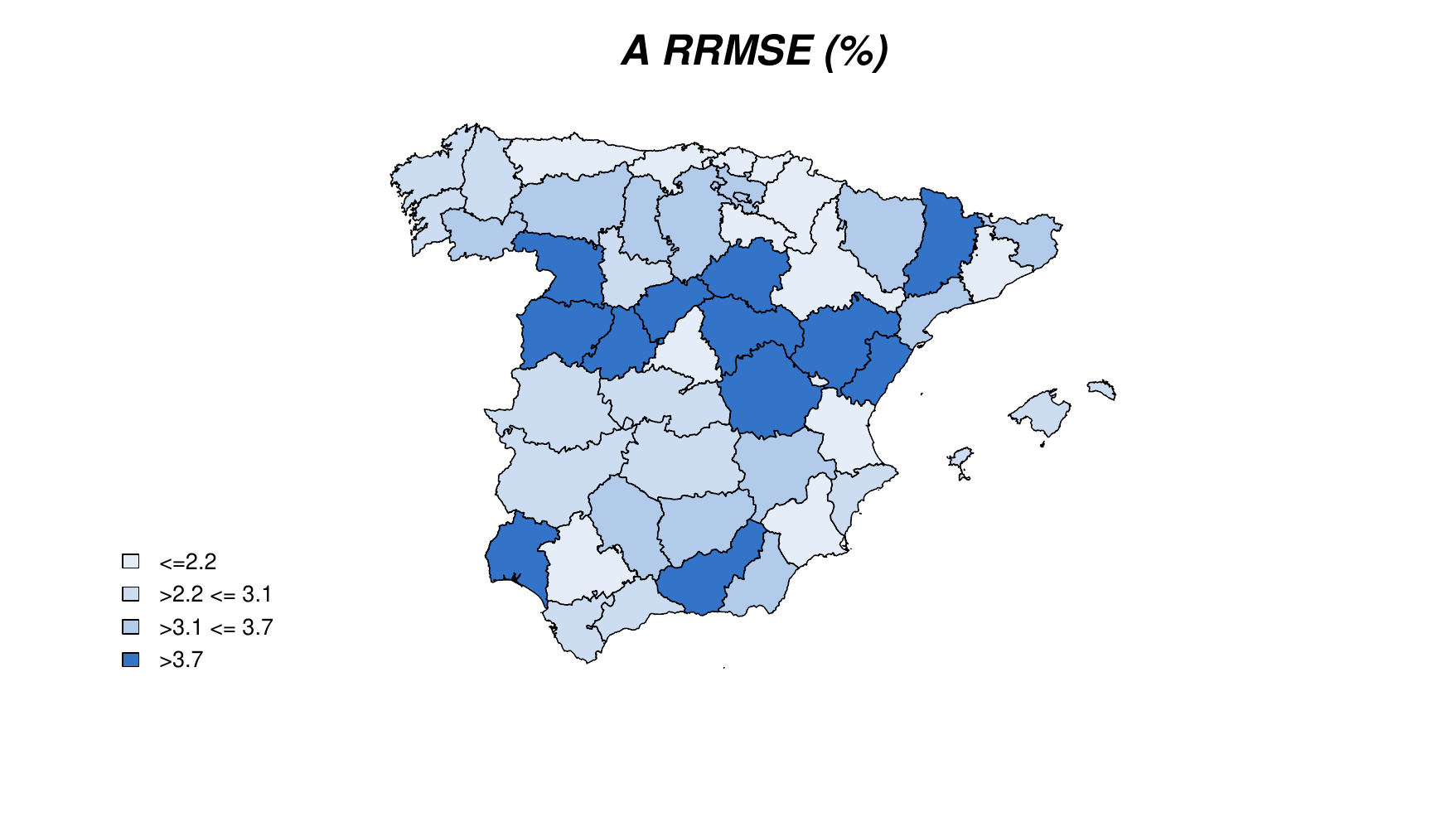}

\vspace*{-3mm}
\caption{Ratios $\hat{A}_{d}$ in \% of household annual expenditures in food by Spanish provinces.}
\label{mapAd}
\end{center}
\end{figure}

Figure \ref{map_percentage_g} (left) and Figure \ref{mapAd} (left) plot the ratios of means and the mean of ratios of household expenditures in food by Spanish provinces (in \%).
Figure \ref{map_percentage_g} (right) and Figure \ref{mapAd} (right) plot the corresponding RRMSEs in \%.
An interesting feature observed here is that within some Autonomous Regions, the province percentages of food expenditure, $R_d$, could be rather variable.
The same happens for the province averages of household percentages of food expenditure $A_d$.
This happens mostly in the Autonomous Regions of Andaluc\'{\i}a, Arag\'on, Castilla Le\'on or in Galicia, where there are many provinces and some of them are more deprived than others.
In contrast, there are other regions, such as Catalu\~na and Basque Country where the variability of the estimated ratios is smaller.

\begin{figure}[!ht]
\begin{center}
\includegraphics[width=130mm,height=60mm]{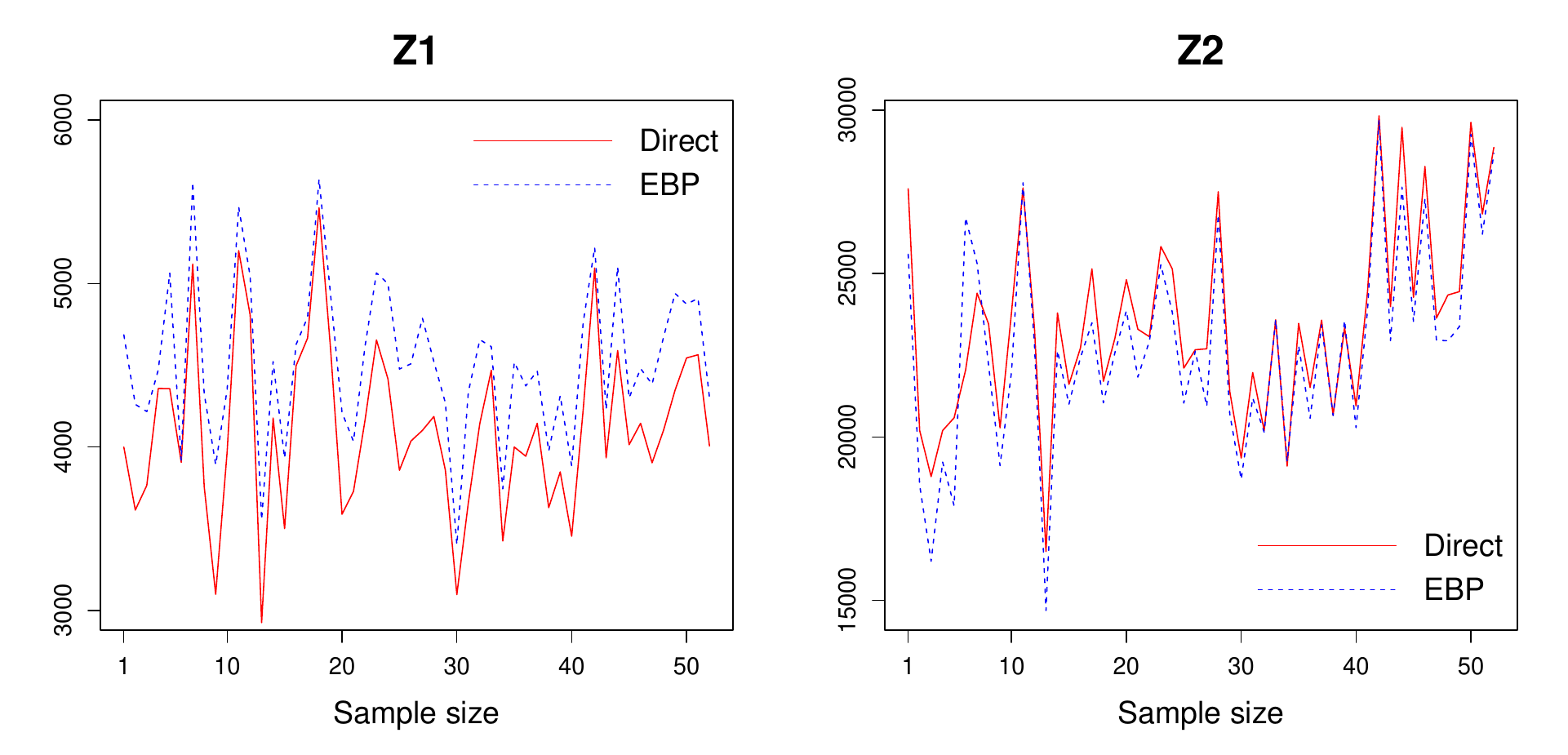}

\vspace*{-3mm}
\caption{Direct and EBP estimates of $\overline{Z}_{d1}$ and $\overline{Z}_{d2}$.}
\label{Y1_Y2_g}
\end{center}
\end{figure}

Figure \ref{Y1_Y2_g} plots the direct and EBP estimates of $\overline{Z}_1$ (left) and  $\overline{Z}_2$ (right).
The domains are sorted by sample sizes and the sample size is printed in the axis OX.
This figure shows that both estimators follow the same pattern and come closer as the sample size increases, but the EBP has a smoother behavior.

\begin{figure}[!ht]
\begin{center}
\includegraphics[width=130mm,height=60mm]{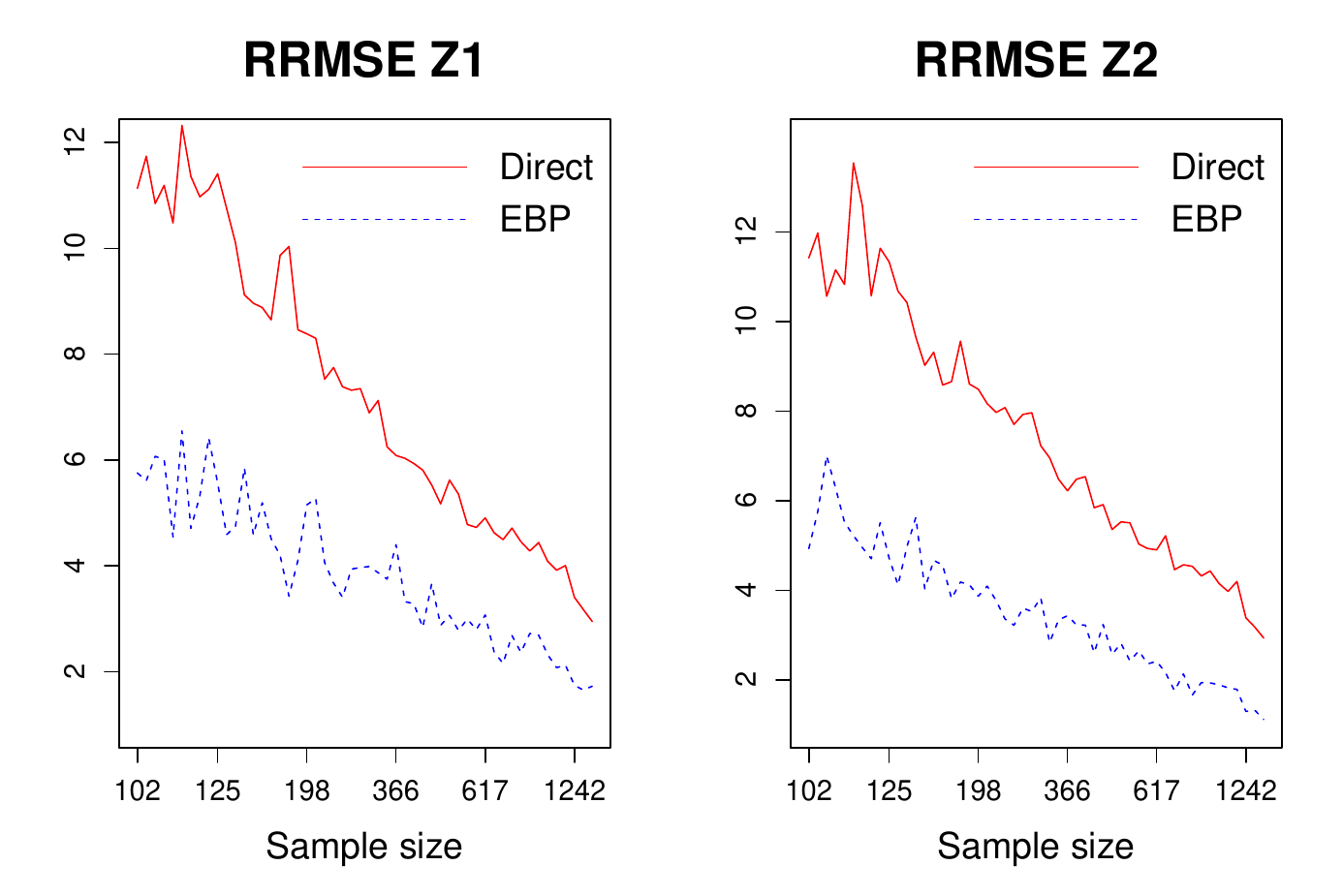}

\vspace*{-3mm}
\caption{RRMSEs of direct and EBP estimates  of $\overline{Z}_{d1}$ and $\overline{Z}_{d2}$.}
\label{rmse_Y1_Y2_g}
\end{center}
\end{figure}

Figure \ref{rmse_Y1_Y2_g} plots the estimated RRMSE of the direct estimators and the EBPs of $\overline{Z}_1$ (left) and  $\overline{Z}_2$ (right).
As before, the domains are sorted by sample size.
This figure shows that the EBPs have lower RRMSEs than the direct estimators and that the RRMSEs come closer as the sample size increases.\\

\vspace*{-3mm}

\begin{figure}[!ht]
\begin{center}
\includegraphics[width=130mm,height=60mm]{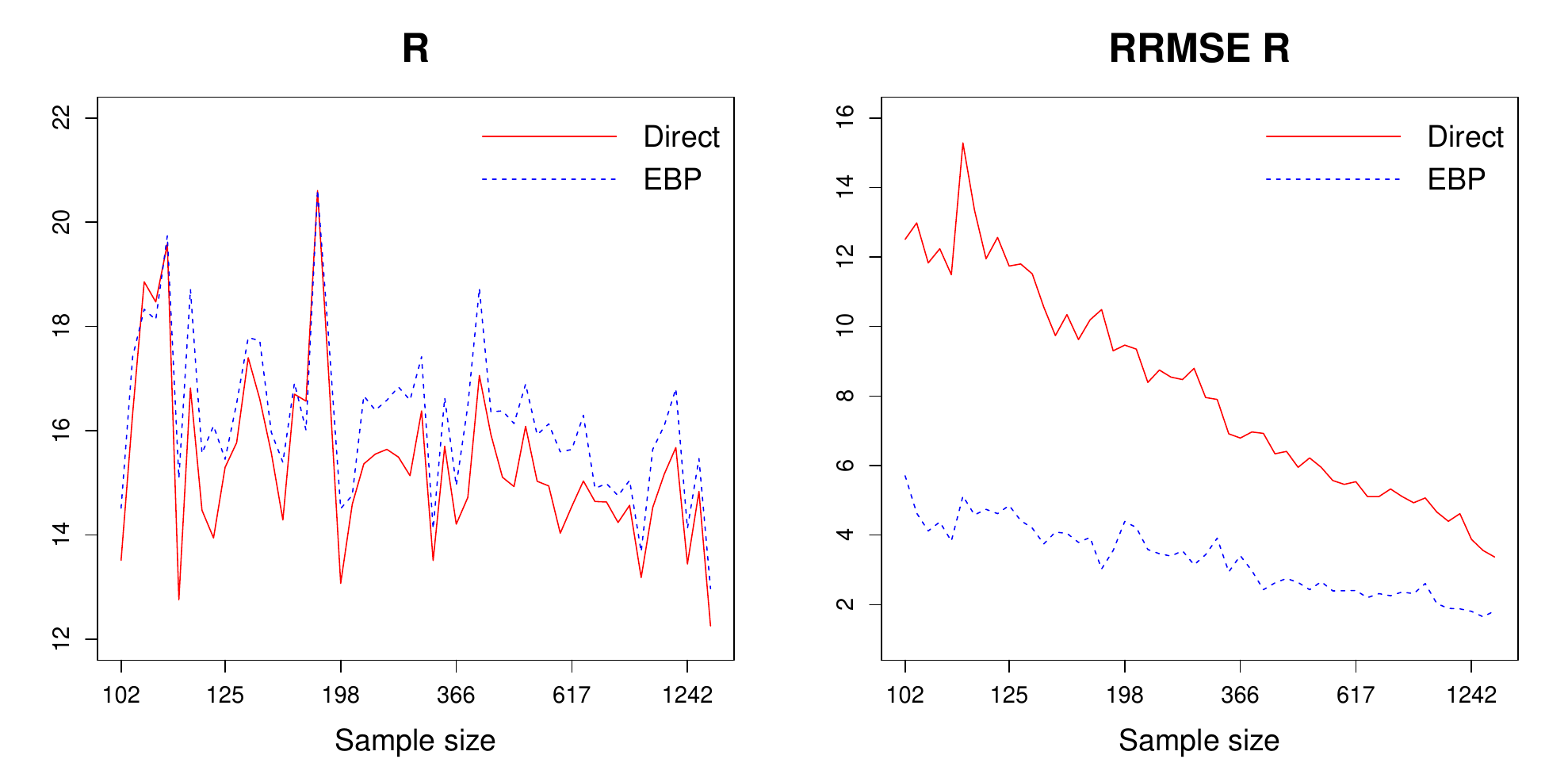}
\caption{Direct, and EBP estimates of ratios $R_d$.}
\label{RRMSE_R_g}
\end{center}
\end{figure}

\begin{figure}[!ht]
\begin{center}
\includegraphics[width=130mm,height=60mm]{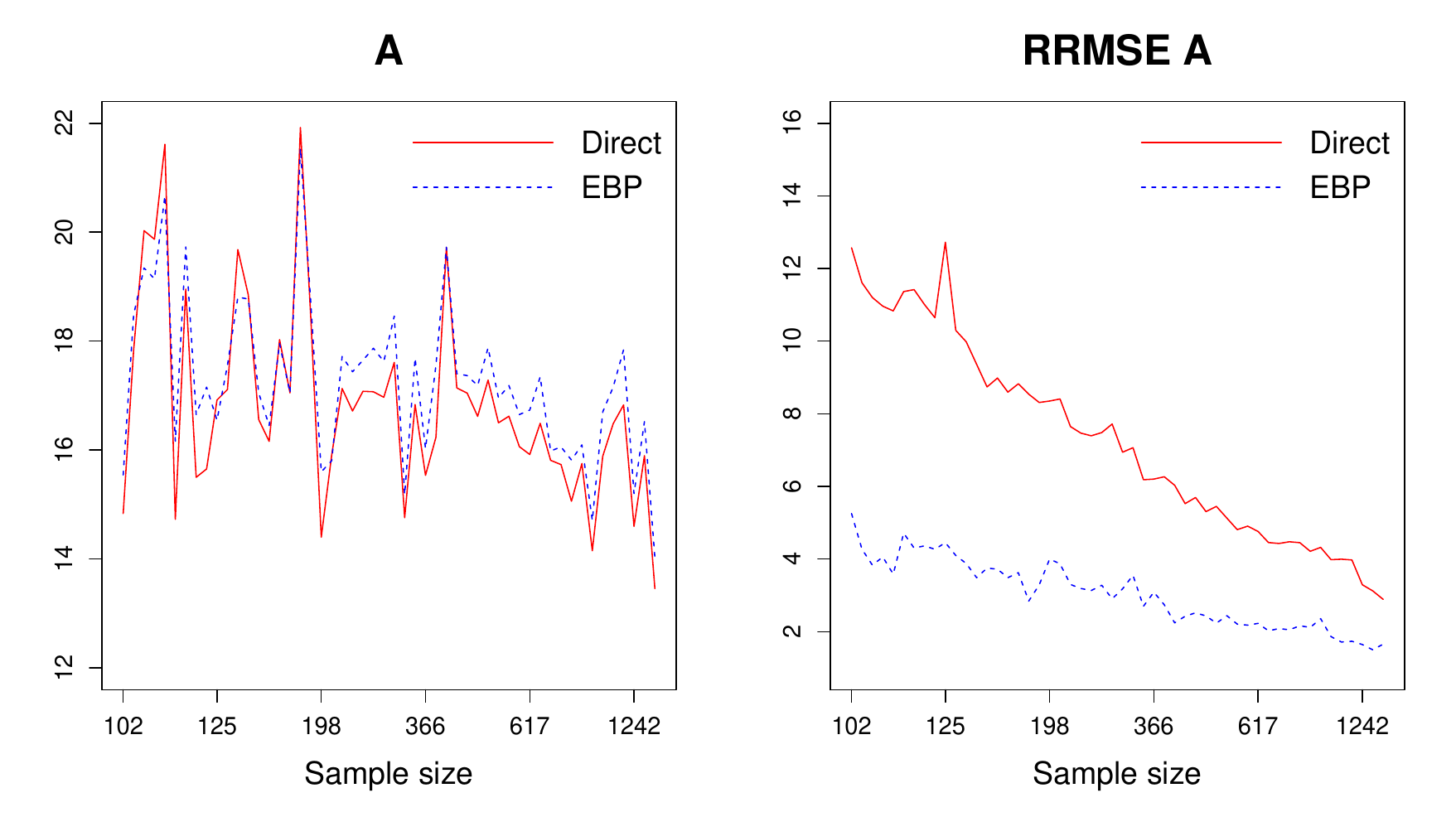}
\caption{Direct, and EBP estimates of ratios $A_d$.}
\label{RRMSE_A_g}
\end{center}
\end{figure}

Figure \ref{RRMSE_R_g} plots the direct and the EBP estimates of province ratios of food expenditure in \% (left) and their corresponding estimated RRMSEs (right).
Figure \ref{RRMSE_A_g} plots the direct and the EBP estimates of the province averages of household ratios of food expenditure in \% (left)
and their corresponding estimated RRMSEs (right).
As before, the domains are sorted by sample size.
These figures show that the model-based estimators have lower RRMSEs than the direct estimators and that the RRMSEs come closer as the sample size increases.

Tables \ref{tabla3} and \ref{tabla4} present some condensed numerical results.
The tables are constructed in two steps.
The domains are sorted by sample size, starting by the domain with the smallest
sample size. A selection of 14 domains out of 52 is done from the positions 1, 5, 9,\ldots,52.
The name and code of provinces are labeled by Prov and $d$ respectively and the sample sizes by $n$.

Table \ref{tabla3} presents the direct and model-based estimates of mean food and non-food household expenditures and the corresponding ratios of food expenditures by provinces.
The estimators are denoted by dir1, ebp1, dir2, ebp2, Rdir, Rebp, Addir, Adebp.
This table shows that the model-based estimates follow the pattern of direct estimates and that both estimates are closer when the sample size is large.

\begin{table}[ht!]
\centering
\begin{tabular}{|lrrrrrrrrrr|}
  \hline
 province& $d$ & $n_d$ & dir1 & ebp1 & dir2 & ebp2 & Rdir & Rebp &Addir&Adebp\\
  \hline
Guadalajara & 19 & 102 & 3999 & 4686 & 25591 & 27596 & 13.52 & 14.52 & 14.83 & 15.54 \\
  Palencia & 34 & 118 & 4357 & 5063 & 17893 & 20587 & 19.58 & 19.74 & 21.61 & 20.67 \\
  Cuenca & 16 & 123 & 3099 & 3891 & 19123 & 20281 & 13.95 & 16.10 & 15.65 & 17.16 \\
  Ourense & 32 & 169 & 2926 & 3555 & 14691 & 16496 & 16.61 & 17.73 & 18.85 & 18.77 \\
  Burgos & 9 & 187 & 4666 & 4793 & 23492 & 25141 & 16.57 & 16.01 & 17.05 & 17.05 \\
  Granada & 18 & 198 & 3729 & 4034 & 21833 & 23296 & 14.59 & 14.76 & 15.91 & 15.80 \\
  Albacete & 2 & 249 & 3858 & 4477 & 21039 & 22107 & 15.49 & 16.84 & 17.07 & 17.87 \\
  Ciudad Real & 13 & 355 & 3858 & 4270 & 20714 & 21396 & 15.70 & 16.64 & 16.84 & 17.69 \\
  Pontevedra & 36 & 463 & 4469 & 4614 & 23593 & 23583 & 15.93 & 16.36 & 17.14 & 17.40 \\
  A Coru\~na & 15 & 536 & 4145 & 4465 & 23429 & 23575 & 15.03 & 15.92 & 16.50 & 16.96 \\
  Zaragoza & 50 & 678 & 4228 & 4761 & 23889 & 24452 & 15.04 & 16.30 & 16.49 & 17.35 \\
  Cantabria & 39 & 761 & 4014 & 4300 & 23536 & 24272 & 14.57 & 15.05 & 15.75 & 16.09 \\
  Murcia & 30 & 913 & 4347 & 4939 & 23379 & 24446 & 15.68 & 16.81 & 16.83 & 17.84 \\
   Madrid & 28 & 1653 & 4006 & 4302 & 28676 & 28859 & 12.26 & 12.97 & 13.46 & 14.03 \\
   \hline
\end{tabular}
\caption{Estimates of $\overline{Z}_{d1}$, $\overline{Z}_{d2}$,  $R_d$ and $A_d$ (in \%).}
\label{tabla3}
\end{table}

\begin{table}[ht!]
\centering
\begin{tabular}{|lrrrrrrrrrr|}
  \hline
  province& $d$ & $n_d$ & dir1 & ebp1 & dir2 & ebp2 &Rdir & Rebp &Addir&Adebp\\
  \hline
Guadalajara & 19 & 102 & 11.13 & 5.75 & 11.42 & 4.93 & 12.51 & 5.71 & 12.57 & 5.25 \\
  Palencia & 34 & 118 & 10.48 & 4.54 & 10.83 & 5.52 & 11.49 & 3.83 & 10.83 & 3.58 \\
  Cuenca & 16 & 123 & 11.11 & 6.42 & 11.64 & 5.51 & 12.56 & 4.62 & 10.65 & 4.27 \\
  Ourense & 32 & 169 & 9.12 & 5.85 & 9.65 & 5.63 & 10.56 & 3.75 & 9.36 & 3.49 \\
  Burgos & 9 & 187 & 9.86 & 4.20 & 8.66 & 3.81 & 10.19 & 3.93 & 8.82 & 3.62 \\
  Granada & 18 & 198 & 8.30 & 5.27 & 8.17 & 4.09 & 9.35 & 4.22 & 8.41 & 3.87 \\
  Albacete & 2 & 249 & 7.32 & 3.94 & 7.93 & 3.60 & 8.47 & 3.55 & 7.48 & 3.28 \\
  Ciudad Real & 13 & 355 & 6.25 & 3.75 & 6.48 & 3.34 & 6.91 & 2.94 & 6.18 & 2.70 \\
  Pontevedra & 36 & 463 & 5.81 & 2.84 & 5.84 & 2.60 & 6.34 & 2.62 & 5.53 & 2.42 \\
  Coru\~na, A & 15 & 536 & 5.35 & 2.78 & 5.50 & 2.42 & 5.95 & 2.66 & 5.13 & 2.44 \\
  Zaragoza & 50 & 678 & 4.62 & 2.37 & 5.22 & 2.16 & 5.11 & 2.20 & 4.45 & 2.02 \\
  Cantabria & 39 & 761 & 4.28 & 2.72 & 4.32 & 1.94 & 4.93 & 2.31 & 4.21 & 2.12 \\
  Murcia & 30 & 913 & 4.00 & 2.12 & 4.19 & 1.78 & 4.62 & 1.88 & 3.97 & 1.73 \\
  Madrid & 28 & 1653 & 2.95 & 1.72 & 2.93 & 1.11 & 3.37 & 1.82 & 2.89 & 1.66 \\
   \hline
\end{tabular}
\caption{RRMSE of $\overline{Z}_{d1}$, $\overline{Z}_{d2}$,  $R_d$ and $A_d$ (in \%).}
\label{tabla4}
\end{table}

\section{Conclusions}\label{sect.con}
This paper introduces small area predictors of expenditure means and ratios based on the BNER model (\ref{modpop2NERdj}).
Best predictors minimize the MSE, within the class of unbiased predictors, under the model distribution.
This optimality property approximately holds if we substitute true model parameters by consistent estimators, as the REML estimators are.
The paper proposes estimating the MSEs of the EBPs by parametric bootstrap.
As far as we know, this is the first time that EBPs for nonlinear bivariate parameters are introduced.

Two simulation experiments are carried out to empirically investigate and to check the behavior of the EBPs and the MSE estimators
when the number of domains or the domain sample sizes are small.
This is to say, in scenarios where the asymptotic properties might not hold.
Simulation 1 investigates the biases and the MSEs of the EBPs.
Simulation 1 shows that the EBPs are basically unbiased, even in cases with rather small sample sizes, and that the MSEs decrease as the sample sizes $n_d$ increase.
Simulation 2 gives the recommendation of doing $B=400$ iterations when applying the introduced parametric bootstrap procedures for estimating the MSEs.

The introduced EBP methodology is applied to data from the SHBS of 2016.
The target is to estimate province means of food and non-food household annual expenditures, ratios of province  means of household annual expenditures and
province means of ratios of household annual expenditures.
The estimation procedure takes into account the correlation between the two target variables.
The paper also compares the model-based estimates with direct estimates and it shows that introduced EBPs have lower MSEs.

%
%
\section*{References}

\begin{description}\setlength\itemsep{-0.1em}
\item
Arima, S., Bell, W.R., Datta, G.S., Franco, C., Liseo, B. (2017).
Multivariate Fay–Herriot Bayesian estimation of small area means under functional measurement error
\textit{Journal of the Royal Statistical Society, series A}, {\bf 180}, 4, 1191-09.
\item
Boubeta, M, Lombard\'{\i}a, M.J., Morales, D. (2016). Empirical best prediction under area-level Poisson mixed models.
\textit{TEST}, {\bf 25}, 548-569.
\item
Boubeta, M, Lombard\'{\i}a, M.J., Morales, D.  (2017). Poisson mixed models for studying the poverty in small areas.
\textit{Computational Statistics and Data Analysis}, {\bf 107}, 32-47.
\item
Benavent, R.,  Morales, D. (2016). Multivariate Fay-Herriot models for small area estimation.
{\it Computational Statistics and Data Analysis}, {\bf 94}, 372-390.
\item[]
Chandra, H., Salvati, N. and Chambers, R. (2017). Small area prediction of counts under a non-stationary spatial model.
{\it Spatial Statistics}, {\bf 20}, 30-56.
\item[]
Chandra, H., and Salvati, N. (2018). Small area estimation for count data under a spatial dependent aggregated level random effects model.
{\it Communications in Statistics - Theory and Methods}, {\bf 47}, 5, 1234-1255.
\item[]
Chandra, H., Salvati, N. and Chambers, R. (2018). Small area estimation under a spatially non-linear model.
{\it Computational Statistics and Data Analysis}, {\bf 126}, 19-38.
\item
Datta, G. S., Fay, R. E., Ghosh, M. (1991).
Hierarchical and empirical Bayes multivariate analysis in small area estimation.
In: Proceedings of Bureau of the Census 1991 Annual Research Conference, U.S. Bureau of the Census, Washington, DC, 63-79.
\item
Datta, G.S., Day, B., Basawa, I. (1999). Empirical best linear unbiased and empirical Bayes prediction in multivariate small area estimation.
\textit{Journal of Statistical Planning and Inference}, {\bf 75}, 269-279.
\item
Esteban, M.D., Lombard\'{\i}a, M.J., L\'opez-Vizca\'{\i}no, E., Morales, D., P\'erez, A. (2020).
Small area estimation of proportions under area-level compositional mixed models. {\it TEST}.  DOI: 10.1007/s11749-019-00688-w.
\item
Esteban, M.D., Lombard\'{\i}a, M.J., L\'opez-Vizca\'{\i}no, E., Morales, D., P\'erez, A. (2021).
Small area estimation of expenditure means and ratios under a unit-level bivariate linear mixed model.
To appear in Journal of Applied Statistics.
\item
Fay, R. E. (1987). Application of multivariate regression of small domain estimation.
In: R. Platek, J. N. K. Rao, C. E. S\"arndal and M. P. Singh (Eds.), Small Area Statistics, Wiley, New York, 91-102.
\item[]
Gonz\'alez-Manteiga, W., Lombard\'{\i}a, M. J., Molina, I., Morales, D., Santamar\'{\i}a, L. (2007).
Estimation of the mean squared error of predictors of small area linear parameters under a logistic mixed model.
{\it Computational Statistics and Data Analysis}, {\bf 51},  2720-2733.
\item[]
Gonz\'alez-Manteiga, W., Lombard\'{\i}a, M. J., Molina, I., Morales, D., Santamar\'{\i}a, L. (2008).
Bootstrap mean squared error of small-area EBLUP.
{\it Journal of Statistical Computation and Simulation}, {\bf 78}, 443-462.
\item
Hobza, T. and Morales, D. (2016). Empirical Best Prediction Under Unit-Level Logit Mixed Models.
\textit{Journal of official statistics}, \textbf{32}, 3, 661-692.
\item
Hobza, T., Morales, D., Santamar\'{\i}a, L. (2018). Small area estimation of poverty proportions under unit-level temporal binomial-logit mixed models.
\textit{TEST}, \textbf{27}, N. 2., 270-294.
\item
Hobza, T., Morales, D., Marhuenda, Y. (2020). Small area estimation of additive parameters under unit-level generalized linear mixed models.
To appear in {\it SORT}.
\item
Ito, T., Kubokawa, T. (2018). Empirical best linear unbiased predictors in multivariate nested-error regression models.
\textit{Communications in Statistics--Theory and Methods}, DOI: 10.1080/03610926.2019.1662048.
\item[]
Jiang, J., P. Lahiri. (2001). Empirical best prediction for small area inference with binary data.
{\it Annals of the Institute of Statistical Mathematics}, {\bf 53}, 217-243.
\item[]
Jiang, J. (2003). Empirical best prediction for small-area inference based on generalized linear mixed models.
{\it Journal of statistical planning and inference}, {\bf 111}, 117-127.
\item[]
L\'opez-Vizca\'{\i}no, E., Lombard\'{\i}a, M.J., Morales, D. (2013).
Multinomial-based small area estimation of labour force indicators. {\it Statistical Modelling}, {\bf 13}, 2, 153-178.
\item[]
L\'opez-Vizca\'{\i}no, E., Lombard\'{\i}a, M.J. and Morales, D. (2015).
Small area estimation of labour force indicators under a multinomial model with correlated time and area effects.
\textit{Journal of the Royal Statistical Association, series A}, {\bf 178}, 3, 535-565.
\item[]
Marchetti, S. and Secondi, L. (2017) Estimates of household consumption expenditure at provincial level in Italy by using small area estimation methods: ``Real'' comparisons using purchasing power parities. \textit{Social Indicators Research}, {\bf 131}, 215-234.
\item[]
Marhuenda, Y., Molina, I., Morales, D., Rao, J.N.K. (2017). Poverty mapping in small areas under a two-fold nested error regression model.
\textit{Journal of the Royal Statistical Society, series A}. {\bf 180}, 4, 1111-1136.
\item[]
Marino, M.F., Ranalli, M.G., Salvati, N., Alfo, M. (2019). Semiparametric empirical best prediction for small area estimation of unemployment indicators.
{\it The Annals of Applied Statistics}, {\bf  13}, 2, 1166-1197.
\item[]
Molina, I., Saei, A. and Lombard\'{\i}a, M. J. (2007) Small area estimates of labour force participation under a
multinomial logit mixed model. \textit{Journal of Royal Statistical Society, Series A}, {\bf 170}, 975-00.
\item[]
Molina, I. and Rao, J.N.K. (2010). Small area estimation of poverty indicators.
{\it The Canadian Journal of Statistics}, {\bf 38}, 369-385.
\item
Molina, I., Mart\'{\i}n, N. (2018). Empirical best prediction under a nested error model with log transformation.
{\it Annals of Statistics}, {\bf 46}, 5, 1961-1993.
\item
Morales, M., Pagliarella, M.C., Salvatore, R. (2015). Small area estimation of poverty indicators under partitioned area-level time models.
\textit{SORT-Statistics and Operations Research Transactions}, {\bf 39}, 1, 19-34.
\item
Ngaruye, I., Nzabanita, J., von Rosen, D., Singull, M. (2017).
Small area estimation under a multivariate linear model for repeated measures data.
\textit{Communications in Statistics - Theory and Methods}, {\bf 46}, 21, 10835-10850.
\item
Porter, A.T., Wikle, C.K., Holan, S.H. (2015). Small Area Estimation via Multivariate Fay-Herriot Models With Latent Spatial Dependence. {\it  Australian \& New Zealand Journal of Statistics}, {\bf 57}, 15-29.
\item
Rojas-Perilla, N., Pannier, S., Schmid, T., Tzavidis, N. (2020). Data-driven transformations in small area estimation.
{\it Journal of the Royal Statistical Society, Series A}, {\bf 183}, 1, 121-148.
\item[]
Torabi, M. (2019). Spatial generalized linear mixed models in small area estimation.
{\it The Canadian Journal of Statistics}, {\bf 47}, 3, 426-437.
\item
Tzavidis, N., Salvati, N., Pratesi, M. and Chambers, R. (2008). M-quantile models with application to poverty mapping.
{\it Statistical Methods and Applications}, {\bf 17}, 3, 393-411.
\item
Ubaidillah, A., Notodiputro, K.A., Kurnia, A., Wayan, I. (2019). Multivariate Fay-Herriot models for small area estimation
with application to household consumption per capita expenditure in Indonesia.
\textit{Journal of Applied Statistics}. Available in \\ https://doi.org/10.1080/02664763.2019.1615420.
\end{description}

\end{document}